# Enhancing Social Media Rumor Detection: A Semantic and Graph Neural Network Approach for the 2024 Global Election

Liu Yan, Liu Yunpeng, Zhao Liang

Department of Computing

The Hong Kong Polytechnic University

Hong Kong

China

# Abstract

The development of social media platforms has revolutionized the speed and manner in which information is disseminated, leading to both beneficial and detrimental effects on society. While these platforms facilitate rapid communication, they also accelerate the spread of rumors and extremist speech, impacting public perception and behavior significantly. This issue is particularly pronounced during election periods, where the influence of social media on election outcomes has become a matter of global concern. With the unprecedented number of elections in 2024, against this backdrop, the election ecosystem has encountered unprecedented challenges.

This study addresses the urgent need for effective rumor detection on social media by proposing a novel method that combines semantic analysis with graph neural networks. We have meticulously collected a dataset from PolitiFact and Twitter, focusing on politically relevant rumors. Our approach involves semantic analysis using a fine-tuned BERT model to vectorize text content and construct a directed graph where tweets and comments are nodes, and interactions are edges.

The core of our method is a graph neural network, SAGEWithEdgeAttention, which extends the GraphSAGE model by incorporating first-order differences as edge attributes and applying an attention mechanism to enhance feature aggregation. This innovative approach allows for the fine-grained analysis of the complex social network structure, improving rumor detection accuracy.

Our experiments on a real-world Twitter dataset demonstrate that the SAGEWithEdgeAttention algorithm outperforms the baseline GraphSAGE model, showcasing the effectiveness of our method in identifying rumors. The integration of BERT's deep contextual understanding with the advanced graph neural network technology provides a robust solution for rumor detection in social networks. The study concludes that our method significantly outperforms traditional content analysis and time-based models, offering a theoretically sound and practically efficient solution.

# Content

# 1. Introduction

With the development of social media platforms such as Twitter and Facebook, information posted by individual users can attract millions of users' attention in a very short period. This not only greatly facilitates people's communication and contact with relatives and friends, but also becomes an important channel for obtaining instant news and understanding current events. Social media platforms have accumulated a huge user base, and trust relationships have been established among their users. These platforms have gradually evolved into the main source of information for many users. However, the quality of information on These platforms has been questioned. After in-depth analysis of Google search results, real-time information spread on platforms such as Twitter and blogs is full of rumors, fabricated content, false lies, misunderstandings, and unconfirmed events (Ahsan, Kumari, & Sharma, 2019).

The "explosive" spread of rumors on social media may cause serious harm to society (Yu, Lu, Wang, & Li, 2021). In the past, rumors spread slowly, but the emergence of Internet technology and the popularity of forwarding activities on social networks have promoted rumors to spread at an alarming rate worldwide. In the 2016 US presidential election, Facebook became the main medium for spreading fake news, which affected people's choice of voting and significantly influenced the election outcomes (Mee, & Vishwakarm, 2020).

Throughout the 2020 United States presidential election period, rumors on social media had a more significant impact on the election than traditional media, distorting voters' perceptions of candidates and voting decisions (Benaissa Pedriza, 2021).

Moreover, researchers, technologists, and political scientists said that false information posed an unprecedented threat to American democracy in 2024. As the US presidential election approached, the academic community issued a warning, pointing out that the complex interweaving of domestic and foreign environments, traditional media, and social media, against the backdrop of authoritarian resurgence, widespread distrust, and intensified political and social instability, the harm of propaganda, lies, and conspiracy theories had reached unprecedented heights, and their danger had significantly increased.

The current year's US presidential election coincides with a critical period of reshaping the global political landscape. Billions of people are participating in various election activities in over 50 nations including Mexico, South Africa, Europe, and India. This historic moment coincides with a period of breeding ground for the spread of false information and its creators. Against this backdrop, the election ecosystem has encountered unprecedented challenges.

Numerous voter groups have shown a high susceptibility to the false information spread by former President Donald Trump and his support camp; on the other hand, the widespread application of artificial intelligence technology has brought convenience, but it has also become a driver for the accelerated spread of rumors and false information; in addition, the efforts of social media platforms to control misinformation within their ecosystem seem to be weakening, further exacerbating the chaos and complexity of the information environment. These phenomena together constitute a severe test for global elections. (NBC News, 2024).

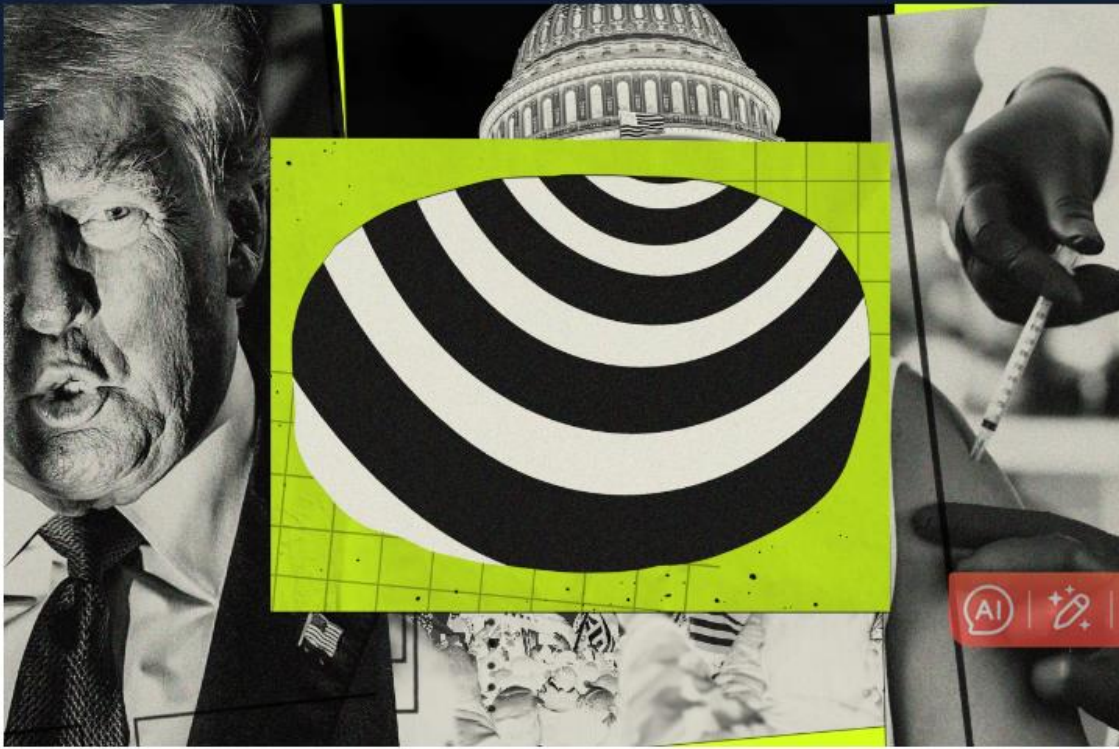

The lie that the 2020 election was "stolen" has proven staggeringly effective with Republicans, with a toll that extends to distrust in future

In the initial stage, researchers mainly relied on manually filtering various attributes of social media platforms, such as user account information, text content, post dissemination patterns, and domain reliability, to advance their research work, and processed these attributes through machine learning methods. Although these methods achieved some success, they were costly and time-consuming. With the development of deep learning technology, researchers began to explore the use of this technology to explore the deep-level features of rumors.

Rumors are a dynamic entity that evolves over time, consisting of a series of posts with timestamps. Therefore, rumors can essentially be viewed as a time series. To capture this temporal characteristic, recursive neural networks (RNNs) are introduced and applied to rumor detection, which can effectively mine the characteristics of posts that change over time. Furthermore, several investigations have endeavored to integrate temporal attributes with user characteristics or image attributes, aiming to enhance the precision of rumor detection processes even further.

However, relying solely on RNNs cannot fully leverage the complex interaction characteristics between rumor posts. To compensate for this drawback, researchers have constructed methods for identifying rumors based on convolutional neural networks. These methods can analyze local neighborhoods in propagation sequences and extract advanced features, but they are still insufficient in capturing global interaction features (Fan, Ma, Li, Wang, Cai, Tang, & Yin, 2020).

Graph Neural Network (GNN) is a deep architecture designed specifically for graph data, aiming to learn meaningful representations from graph data by leveraging the graph structure and node features itself. The fundamental concept revolves around iteratively amalgamating feature data from the immediate graph vicinity of each node, leveraging the deep neural networks for this integration process. In this process, node information undergoes transformation and aggregation operations, allowing it to spread widely along the graph structure. Therefore, GNN naturally integrates node features and topological structure information (Han, Yu, Su, & Wu, 2023). Given the rich interaction relationships between rumor posts, researchers have constructed an interactive graph of rumors and introduced GNN. GNN can efficiently capture the global interaction features in the graph, thereby offering a novel and insightful approach to the task of rumor detection. In the latest research, GNN has proven their efficacy in the realm of rumor detection, achieving notable and substantial results that demonstrate their value in this domain. (Fan, Ma, Li, Wang, Cai, Tang, & Yin, 2020).

Therefore, In response to address the challenge of rumor detection on social media, we propose a practical and effective method. Our method makes the following key contributions:

1. Data Collection for a Politically Relevant Rumor Dataset

To build a robust dataset for rumor detection, we meticulously collected data from two primary sources, PolitiFact and Twitter (now X platform). From PolitiFact, we extracted a news event veracity judgment dataset, which includes crucial information such as the judgment label, statement content, statement source platform, statement date, and a list of news sources. Additionally, we specifically crawled Twitter to gather original tweets related to news events. This rich dataset encompasses original tweets, comments and retweets, replies to comments, and associated user information.

2. Semantic Analysis of Social Media Data

To delve into the collected data, we employed semantic learning techniques. Utilizing a fine-tuned language model, we represented the tweets and comments corresponding to each node in the graph. Specifically, we used the BERT (Bidirectional Encoder Representations from Transformers) model to vectorize the text content. Following vectorization, we constructed a directed graph where main tweets, retweets, and comments are defined as nodes, and comment and retweet actions serve as edges.

3. Rumor Detection Based on Graph Neural Networks

Building upon GraphSage base on GNN, we incorporated first-order differences between nodes as edge attributes. These attributes were then passed through an Attention algorithm and concatenated to the aggregated node features. This processing enabled us to accurately classify the main tweet as rumor or not based on its category.

# 2. Related Work

Graph Neural Networks (GNNs) have become pivotal in the detection of rumors and fake news due to their capability to process and analyze complex graph-structured data. GNNs excel at capturing the intricate relationships and patterns within social networks, where information propagates dynamically. Their evolution from basic graph convolutional models to advanced architectures like GAT and RGCN has significantly enhanced their ability to generalize and adapt to various graph sizes and structures, making them ideal for modeling the temporal dynamics of news diffusion.

The incorporation of temporal elements into GNNs, as seen in models like the Temporally Evolving Graph

Neural Network (TGNF), acknowledges the continuous change in social media interactions. These models consider the sequence of information spread and user engagement over time, offering a more accurate representation of the news propagation process. Concurrently, the fine-tuning of pre-trained language models like BERT has fortified textual analysis in rumor detection, enabling a deeper understanding of content veracity through context-aware representations.

In conclusion, the synergy of advanced GNNs and fine-tuned language models presents a robust approach to fake news detection. This integration leverages the collective strengths of structural graph analysis and nuanced textual comprehension, paving the way for more effective and accurate identification of misinformation in the ever-changing landscape of social media.

## 2.1. Graph Neural Networks

The realm of rumor detection within social media platforms has garnered considerable research attention, with a plethora of methodologies being devised to address this intricate challenge. This comprehensive literature review endeavors to offer a cohesive overview of these approaches, with a particular emphasis on their classification strategies, encompassing content-based, structure-oriented, and time-series methods. By examining these various methodologies, we aim to provide a holistic understanding of the current landscape in rumor detection research.

### 2.1.1. Content-Based Classification

Content-based approaches primarily focus on analyzing textual or visual features within social media posts. For instance, Jin et al. (2017) developed an innovative RNN-based model for detecting multimodal fake news, integrating visual and textual elements through an attention mechanism. Similarly, Wang et al. (2018) proposed a multi-task learning framework aimed at learning transferable textual and visual feature representations across diverse posts. Additionally, Zhou et al. (2020) introduced an advanced fake news detection technique that considers complex relationships across different modalities. These methods leverage semantic information and have demonstrated promising results, despite facing challenges such as the inherent difficulty in distinguishing fake news from genuine news based solely on content (Shu et al., 2020).

### 2.1.2. Structure-Based Classification

Structure-based methods leverage interactions among social media users or content, such as commenting on tweets, retweeting, and following relationships, in the propagation of news. These methods can be categorized into two main groups: isomorphic graph-based and heterogeneous graph-based approaches.

#### 2.1.2.1. Isomorphic Graph-Based Classification

Isomorphic graph-based approaches focus on homogeneous networks that consist of a single type of node and edge. For instance, Vosoughi et al. (2018) investigated the spread of false and true news on Twitter using a homogeneous graph framework. Similarly, Ma et al. (2018b) employed a tree-structured RNN, utilizing both top-down and bottom-up approaches, to integrate textual features with propagation structure characteristics. Bian et al. (2020) introduced a bidirectional Graph Convolutional Network (GCN) to learn representations of content semantics and news diffusion graphs. Additionally, Xu et al. (2022) applied GCNs and Gated Graph Neural Networks (GNNs) to capture high-level text content representations and update word-level features, respectively, thereby enhancing rumor detection accuracy through hierarchical feature aggregation.

#### 2.1.2.2. Heterogeneous Graph-Based Classification

Heterogeneous graph-based approaches are designed to manage networks that encompass multiple types of nodes or edges. For example, Yuan et al. (2019) encoded local semantic and global structural information of diffusion graphs using a heterogeneous graph consisting of posts, comments, and users. Huang et al. (2020) introduced a meta-path-based heterogeneous Graph Attention Network (GAT) framework to capture global semantic relationships in text content, breaking down the heterogeneous graph into subgraphs for tweets-words and tweets-users. Zhao et al. (2020) developed a semi-supervised graph embedding model based on GAT to detect spam bots in social networks, capturing diverse neighborhood relationships among users in directed social graphs. Yang et al. (2020) utilized a heterogeneous information network to model rich entity information and employed a graph adversarial learning framework to enhance learning of distinctive structural features. Lin et al. (2020) proposed ClaHi-GAT for rumor detection on Twitter, representing conversation threads as undirected interaction graphs to capture bidirectional patterns between tweets. Huang et al. (2019) leveraged GCNs to model users, considering content, users, and propagation as critical aspects of rumor detection. Shakshi et al. (2021) applied GCNs to capture network properties in the Twitter user-reply graph. Bhavtosh et al. (2021) and Wu et al. (2020) introduced attention-based GNN models incorporating trust and credibility features from a user's neighborhood, integrating historical behavioral data and network structure. Additionally, Wu et al. (2020) presented a propagation GNN with an attention mechanism focusing on the replies-to relationship. Gao et al. (2022) designed a heterogeneous GNN to predict users' future behaviors in social networks, considering both user and event nodes.

### 2.1.3. Time-Series-Based Classification

Time-series-based methods focus on understanding the temporal dynamics of networks, often utilizing discrete-time dynamic graphs (DTDG) or continuous-time dynamic graphs (CTDG). DTDG approaches aim to derive node embeddings by aggregating information from snapshots of the graph (Lu et al., 2019; Manessi et al., 2020; Sankar et al., 2020). In contrast, CTDG approaches emphasize capturing the temporal evolution of the network and dynamically learning node embeddings in continuous time (Kumar et al., 2019; Trivedi et al., 2019; Zhang et al., 2020). Song et al. (2022) introduced a Dynamic Graph Neural Network (DGN) that models a sequence of time snapshots in news dissemination graphs to capture the evolving dynamics of information spread. Temporal Graph Networks (TGNs) provide a framework for deep learning on dynamic graphs represented as sequences of timed events (Rossi et al., 2020). Song et al. (2021) developed a Temporally Evolving Graph Neural Network (TGNF) for detecting fake news on social media, employing temporal Graph Attention Networks (TGAT) to capture dynamic structures, content semantics, and temporal information in news propagation.

### 2.1.4. Hybrid Methods

Hybrid methods integrate both structural and content-based information to ensure that no valuable features are overlooked. Zhong et al. (2020) exemplified this by introducing a Graph Convolutional Network (GCN)-based technique that combines insights derived from the comment-tree structure with the textual content of posts and their associated comments.

Song et al. (2023) proposes a rumor detection model based on GNN, which combines temporal and interactive features. Benslimane et al. (2023) presents a GNN based approach for controversy detection on social media, specifically Reddit. The approach leverages both the graph structure of user interactions and textual content features. To detect early rumor on social media, Huang et al., (2023) developed the SBAG model which consists of two main components, Social Bot Detection (SD), which is a pre-trained multi-layer perception network that captures features to compute the bot possibility for each user and Bot-Aware Graph Rumor Detection (BAG), a GNN that incorporates the bot possibility scores from SD.

## 2.2. Conclusion

Detecting rumors on social networks using Graph Neural Networks (GNNs) is a multifaceted task that necessitates integrating diverse methodologies. The advancement of this field hinges on developing sophisticated models capable of synergistically combining content-based, structure-based, and time-series-based approaches to enhance accuracy and resilience in rumor detection. Recent research trends highlight a shift towards hybrid models that amalgamate various types of data, encompassing user behavior, network topology, and temporal dynamics. Additionally, attention mechanisms have gained prominence for their role in prioritizing pertinent features or nodes within these models. Moreover, there is a burgeoning interest in constructing models capable of providing interpretability for their predictions, crucial for deciphering the underlying mechanisms of rumor propagation and formulating effective mitigation strategies.

## 2.3. Fine-Tuning of Pre-trained Language Model

The field of rumor detection and fake news identification encompasses a diverse array of strategies and methodologies. Early foundational studies, such as those by Kumar et al. (2018) and Shin et al. (2018), provided initial insights into the characteristics of fake news and theoretical frameworks for rumor detection. Hybrid approaches, exemplified by Bondielli et al. (2019) and Ahmed et al. (2017), underscored the effectiveness of integrating different feature sets to enhance detection rates.

Impactful studies, like that of Allcott et al. (2017), quantified the influence of fake news during pivotal events such as the 2016 U.S. Presidential Election, highlighting the critical importance of research in this area. Data mining and machine learning techniques have been extensively applied, with researchers such as Shu et al. (2019) and Zhou et al. (2018) focusing on linguistic pattern analysis and network behavior to detect rumors.

The advent of deep learning models, particularly Convolutional Neural Networks (CNNs) and Long Short-Term Memory networks (LSTMs), as demonstrated by Liu et al. (2018) and O'Brien et al. (2018), has significantly advanced the field. BERT, a state-of-the-art NLP model, has also shown promise in rumor detection tasks, as evidenced by studies like Jwa et al. (2019), which utilized BERT to analyze textual relationships for fake news detection.

Research has also explored BERT in combination with other models. For instance, Wang et al. (2017) proposed a hybrid architecture involving CNNs and LSTMs for fake news detection. Comparative studies, such as those conducted by Ruchansky et al. (2017) and Singh et al. (2017), have further refined our understanding of how linguistic features and user behaviors can be leveraged to identify fake news.

In conclusion, the literature provides a comprehensive overview of the multifaceted approaches to rumor detection in social media. The fine-tuning of BERT, in particular, offers a promising avenue for research, building upon extensive prior work in the field. By synthesizing insights and methodologies from these studies, our research aims to contribute to the development of more accurate and robust systems for detecting rumors and fake news in the dynamic landscape of social media.

# 3. Problem Formulation

In this section, we begin by outlining key definitions and symbolic markers essential for understanding this paper. Next, based on the "who-replies-to-whom" relationship, we delve into the construction of the propagation graph. Finally, we formally present the problem statement for the rumor detection task.

When a user posts a tweet on Twitter, other users can interact with it through sharing, replying, commenting, or retweeting, leading to widespread dissemination of the original post's information across the Internet. To facilitate a clearer comprehension of the propagation structures of posts, we define some terms and symbols used throughout this paper as follows:

**Definition 1** (Source tweet). A source tweet is an original tweet that neither replies to nor retweets any other tweets. We denote the source tweet as $s^{(j)}$, with the superscript $j$ representing its index.

**Definition 2** (Responsive tweet). A responsive tweet is defined as a tweet that replies to the source tweet or other responsive tweets, or one that retweets other tweets. We use $x_i^{(j)}$ to represent the i-th responsive tweet by time which is relevant to the source tweet $s^{(j)}$.

**Definition 3** (Post set). Each post set $C^{(i)}$ consists of a source tweet and all its responsive tweets, for example, $C^{(j)} = \{s^{(j)}, x_1^{(j)}, x_2^{(j)}, \ldots, x_i^{(j)}\}$. To unify the symbol, the source tweet $s^{(j)}$ in post set $C^{(j)}$ can also be represented as $x_0^{(j)}$. So, we defined the Twitter rumor detection dataset as $C = \{C^{(1)}, C^{(2)}, \ldots, C^{(K)}\}$, which indicates the graphs of different source tweets.

Based on reply relationships, each set of posts can form a propagation tree structure (Wu et al., 2015; Ma et al., 2017). To expand the relationships among tweets, we incorporate retweet connections. Specifically, when a tweet retweets the source tweet, it becomes a leaf node in the structure of the source tweet. In our research, we extend the original propagation tree concept to a graph structure. This graph consists of multiple nodes and two distinct types of relational paths. Node $v_i^{(j)}$ in graph $G^{(j)}$ corresponds to tweet $x_i^{(j)}$ in the post set $C^{(j)}$, and node set $V^{(j)} = \{v_0^{(j)}, v_1^{(j)}, \ldots, v_i^{(j)}\}$ corresponds to all nodes of graph $G^{(j)}$.

**Definition 4** (Relation path). For tweets $x_i^{(j)}$ and $x_{i'}^{(j)} (i < i')$, if a reply or retweet relationship exists between two nodes, then the static graph $G^{(i)}$ includes a directed relational path from node $v_i^{(j)}$ to node $v_{i'}^{(j)}$, and we use $e_n^{(j)}$ to represent the edge of node $v_i^{(j)}$ to node $v_{i'}^{(j)}$. We define edge set $E^{(j)} = \{e_1^{(j)}, e_2^{(j)}, \ldots, e_n^{(j)}\}$ as to all edges of graph $G^{(j)}$. As the edges have direction, we define the responsive tweet as the source node of an edge, and define the tweet being responded to as the target node of an edge.

**Definition 5** (Static graph). With the definition of relation path, we can extend the homogeneous tree structure into a homogeneous graph, represented as $G^{(j)} = (V^{(j)}, E^{(j)})$, and $V^{(j)}$ represents the set of all nodes in the propagation graph, with each node $v_i^{(j)}$ corresponds to tweet $x_i^{(j)}$ in the post set $C^{(j)}$. $E^{(j)}$ denotes the set of all edges in the propagation graph, where each edge $e_n^{(j)}$ corresponds to the relationship between two tweets.

**Definition 6** (Dynamic graph). Similar to static graph, a continuous dynamic news propagation graph $G^{(j,t)} = (V^{(j,t)}, E^{(j,t)})$ consists of node set $V^{(j,t)} = \{v_0^{(j,t_1)}, v_1^{(j,t_2)}, \ldots v_{N_{(t)}}^{(j,t_{N_{(t)}})}\}$ and edges set $E^{(j,t)} = \{e_1^{(j,t_1)}, e_2^{(j,t_2)}, \ldots e_{M_{(t)}}^{(j,t_{N_{(t)}})}\}$ at time t. Each node $v_i^{(j,t_i)} \in V^{(j,t)}$ indicates that the tweet $c_i^{(j)}$ which is published at time $t_i$, and $V^{(j,t)}$ represents the nodes belonged to $G^{(j,t)}$ at time $t$. Each edge $e_n^{(j,t_n)} \in E^{(j,t)}$ means that node $v_p^{(j,t_n)}$ has a response to $v_q^{(j,t_i)}$ at time point $t_n$. Specifically, to learn the temporal representation of each node, the response behavior (i.e., $e_n^{(j,t_n)}$) is modeled as an interaction event between tweet $c_p^{(j)}$ and tweet $c_q^{(j)}$. $N_{(t)} = |V^{(t)}|$ is the total number of tweets at time $t$. Similarly,

$M_{(t)} = |E^{(t)}|$ is the total number of interaction events (i.e., reply or retweet) at time point $t$.

After the propagation graph is constructed, we defined the Twitter rumor detection dataset as a collection of static graphs $G = \{G^{(1)}, G^{(2)}, \ldots, G^{(N)}\}$. Further, we can obtain the graph structure of each static graph at multiple time points, thereby obtaining multiple dynamic graphs of the static graph in the time dimension, and use the following formula to represent it, $G = \{G^{(1,t_1)}, \ldots, G^{(1,t_\alpha)}, G^{(2,t_1)}, \ldots, G^{(2,t_\beta)}, \ldots, G^{(N,t_1)}, \ldots, G^{(N,t_\gamma)}\}$.

Rumor detection is typically framed as a multi-class classification task within the context of supervised learning. In this paper, we classify source tweets into five distinct categories, "False", "Pants on Fire", "Half True", "Mostly False" or "Mostly True". Our goal is to learn a classifier from the labeled propagation graph set, that is f, $G^{(j)} \rightarrow Y^{(j)}$, where $Y^{(j)}$ takes one of five classes. For source tweet $s^{(j)}$, the classifier $f$ can figure out the classification result of propagation graph $G^{(j)}$.

# 4. Datasets

To verify the effectiveness of the rumor identification algorithm in this study, we constructed a real-world dataset of politically related rumors through data collection. The collection scope includes the PolitiFact website and Twitter.

## 4.1. Introduction to the PolitiFact

PolitiFact is a nonpartisan, nonprofit fact-checking website operated by the Poynter Institute in Tampa. Its mission is to reduce false information and misleading statements in political speech and provide the public with a clear and accurate political information environment through professional fact-checking work.

The main task of this website is to conduct thorough fact-checking on political figures, political advertisements, political news reports, etc., and assign ratings such as "False", "Pants on Fire", "Half True", "Mostly False" or "Mostly True". The fact-checking team of PolitiFact consists of experienced journalists and editors who use their professional news literacy and fact-checking skills to strictly verify and evaluate each piece of information.

In addition, PolitiFact also pays attention to political speech on social media and responds quickly to widely circulated rumors and false information to dispel them. Through collaboration with major social media platforms, it directly pushes the results of fact-checking to users, helping the public identify and avoid being misled.

The PolitiFact organization utilizes a unique rating system to provide a quantitative assessment of the credibility of the verified subjects. The Truth-O-Meter rating system is divided into five categories, "True," indicating that the statement is accurate with no significant omissions; "Mostly True," indicating that the statement is basically correct but may require further clarification or supplementation of additional information; "Half True," indicating that the statement is only partially accurate with important details missing or a biased understanding of the context; "Mostly False," indicating that the statement contains true elements but leaves the reader with a misleading impression due to the omission of key facts; "False," indicating that the statement contains significant inaccuracies; and "Pants on Fire," indicating that the statement is not only inaccurate but also contains absurd and false elements.

## 4.2. Introduction to Twitter

Each fact-checking news report on the PolitiFact website includes the source platform of the news, such as Instagram posts, TikTok posts, Facebook posts, X posts, Viral image, etc. Currently, we have analyzed the fact-checking news reports whose source platform is X posts and filtered out the links to the news sources on the X platform or Twitter. Then, we collected data from the links on the X platform or Twitter, including the fact-checking news reports.

Twitter, founded by Jack Dorsey and his partners in March 2006 and officially launched in July of the same year, is a trendsetting social media platform that allows users to post messages of up to 140 characters (with an expanded limit of 280 characters for Chinese, Japanese, and Korean). These messages are known as "tweets." On July 23, 2023, CEO Musk announced through a tweet that the iconic "bluebird" icon would be replaced with an "X" icon, marking a significant update to the brand image. Subsequently, on July 24, Musk revealed that Twitter's website address would migrate from twitter.com to X.com. On July 31, Twitter's app name on the Apple App Store was officially changed to X.

## 4.3. PolitiFact Data Collection and Filtering Methodology

In this study, Python scraping technology was utilized to systematically collect all fact-checking news reports published by PolitiFact since its establishment in 2007, totaling approximately 24,000 records. The specific collection process is outlined below,

### 4.3.1. Scraping PolitiFact

Determining the maximum page range, By analyzing the URL structure of the PolitiFact website's fact-checking news report list page (https://www.politifact.com/factchecks/?page=XXX), a large page number (such as 900) was input and observed for the "No Results found" prompt. Gradually decrementing the page number until the current valid page range was determined to be 809 pages.

Iterating and scraping pages: From page 1 to page 809, each page was parsed and links to fact-checking news reports were extracted. Subsequently, the corresponding news report pages were downloaded for each link.

Parsing news report content: The downloaded content of each fact-checking news report page was parsed to extract information including the judgment result, statement, statement source platform, statement date, factchecker name, factcheck date, topic, page number, factcheck analysis link, data acquisition date, and list of news sources.

Data recording and storage: Each fact-checking news report was assigned a unique ID and page number, then organized into a file. When initiating HTTP requests, a random waiting time mechanism was introduced to ensure friendly access to the original website and avoid unnecessary access pressure.

Data conversion and storage: After completing the scraping of all pages, the raw data was converted into JSON and parquet formats, and further transformed into XLS file format for subsequent data processing and analysis.

### 4.3.2. Filtering PolitiFact

Platform source filtering: Within the XLS file, all fact-checking news reports originating from twitter or X platforms were filtered based on the statement source platform.

Analysis of news sources: For the filtered fact-checking news reports, a deep analysis was conducted on the links in the "news source list," with a particular focus on archive website links such as https://ghostarchive.org

and https://archive.is.

Link translation and attribute addition: Using scraping technology, archive website links were converted to their corresponding original links. Subsequently, a new attribute was added to each fact-checking news report, documenting the translated news source address list, labeled as "translated link news source address."

Specific domain filtering: Finally, the "translated link news source address" was further filtered to retain links with root domains of twitter.com, X.com, and m.twitter.com. This resulted in obtaining the original news links for all fact-checking news reports originating on twitter or X platforms.

## 4.4. Twitter Data Collection and Filtering Methodology

The underlying data acquisition on the web version of Twitter primarily relies on the GraphQL interface, a data query and manipulation API that provides developers with a flexible and efficient way to access and manipulate content on the Twitter platform. As a declarative query language, GraphQL allows developers to construct customized requests to precisely retrieve, update, or delete specific content on Twitter.

For this study, the Python library 'twitter-api-client' was employed to automate the scraping of news source links related to fact-checking news reports published on Twitter or X platform using the GraphQL interface.

The GraphQL interface return a JSON format data; after parsing the JSON data, we can obtain Twitter's data. The JSON format data example is shown in Fig. 1, as follows,

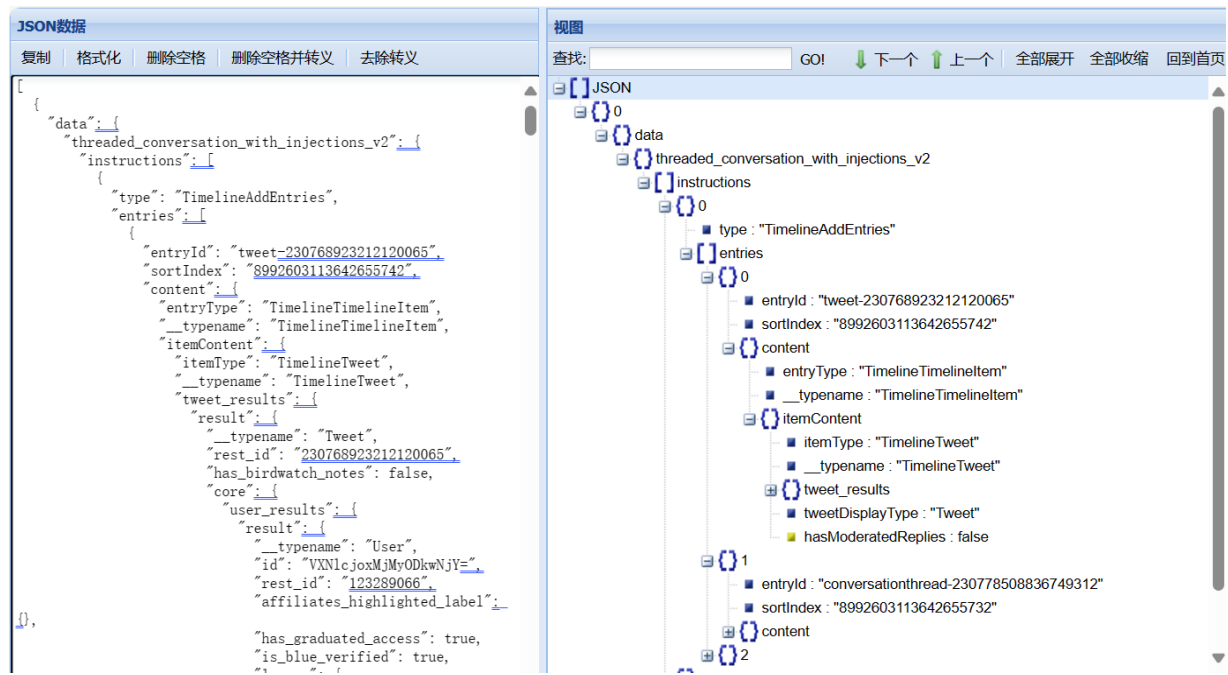

Fig. 1.

The data collected mainly encompassed the following four aspects,

User Information: Detailed user information was scraped for the primary user who posted each tweet, including username, user description, number of followers, and following list.

Tweet Information: Key metrics such as tweet content, publication time, retweet count, and like count were extracted. These data points are crucial for analyzing the tweet's dissemination effectiveness and audience feedback.

Retweet and Comment Information: In addition to the primary tweet, all retweets and their respective user information were scraped. Furthermore, all comments and their authors' user information were collected from under each tweet. This information helps gain a deeper understanding of the tweet's propagation path and audience interaction patterns.

Secondary Comment Information: To comprehensively capture the dynamic dissemination of tweets, secondary comments were also scraped, including all replies to original comments and retweets along with the respective user information.

Through these steps, a multi-layered dataset encompassing "fact-checking news reports - primary tweets - comments/retweets - replies to comments/retweets" was constructed. Simultaneously, detailed information of all relevant users was collected. This dataset not only provides abundant material for subsequent text analysis and network analysis but also serves as a powerful data support for studying the dissemination mechanisms and audience interaction patterns of fact-checking news reports on the Twitter.

## 4.5. Datasets Detail
### 4.5.1. PolitiFact Data

We have comprehensively obtained all factcheck items from PolitiFact since its establishment in 2007, totaling 23,930 articles. After filtering and verifying all data labeled as X posts in the news reports, a total of 166 verified factcheck items were obtained. Among these 166 verified factcheck items, 16 of the source links in the data do not include Twitter.com or x.com, so they are excluded. Finally, we obtained a dataset of 150 verified factcheck items from the Twitter platform. The table, Table 1, and field explanations for PolitiFact data are as follows,

Table 1

| NO. | Field | Field explanations |
| --- | --- | --- |
| 1 | verdict | Judgment result |
| 2 | statement | Statement |
| 3 | statement_originator | Platform of statement origin |
| 4 | statement_date | Date of statement |
| 5 | factchecker_name | Name of factchecker |
| 6 | factcheck_date | Date of factcheck |
| 7 | topics | Topics |
| 8 | page | Page number |
| 9 | factcheck_analysis_link | Link to factcheck analysis |
| 10 | date_retrieved | Date of data retrieval |
| 11 | oursource_links | List of source links |
| 12 | translate_links | Source link address after translation |
| 13 | translate_twitter_links | Source link with tweet address after translation |

The PolitiFact website webpage corresponding to the PolitiFact data field is as follows in Fig. 2 and Fig. 3,

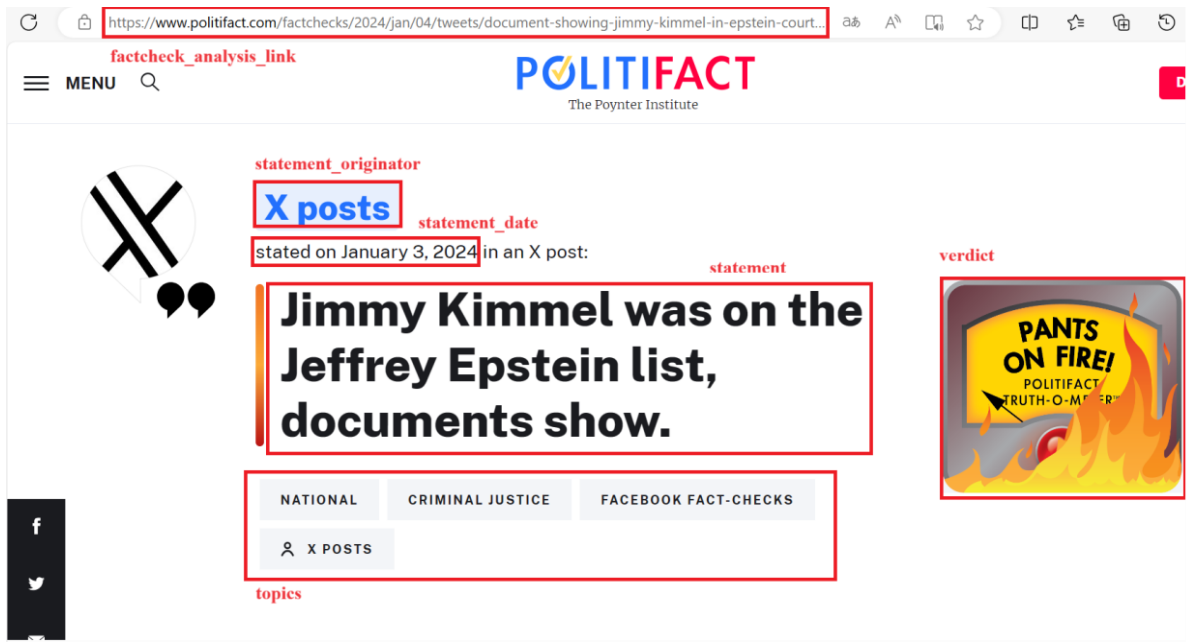

Fig. 2.

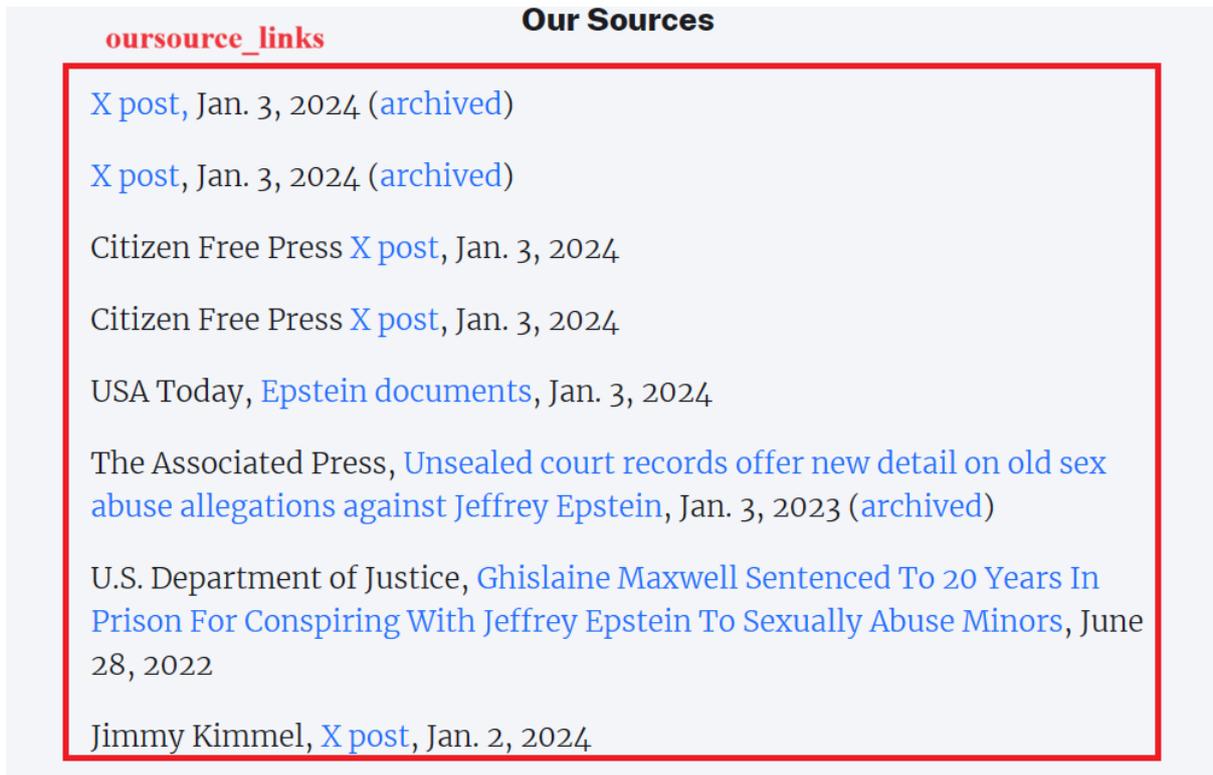

Fig. 3.

### 4.5.2. Twitter Data

According to the Twitter collection method above, we collected a total of 94,440 tweet data, 94,440 comment data, 618,387 retweet data, and 523,638 user data.

### 4.5.2.1. Tweet data

The table and field explanations for Twitter tweet data are as follows in Table 2,

Table 2

| NO. | Field | Field explanations |
| --- | --- | --- |
| 1 | id | Tweet ID |
| 2 | link | Original tweet link |
| 3 | date | Tweet date |
| 4 | user_id | User ID |
| 5 | username | Username |
| 6 | place | Location |
| 7 | tweet | Tweet content |
| 8 | mentions | Mentioned users |
| 9 | replies | Number of replies |
| 10 | retweets | Number of retweets |
| 11 | hashtags | Hashtags |
| 12 | cashtags | Cashtags |
| 13 | language | Language |
| 14 | place_code | Place code |
| 15 | place_id | Place ID |
| 16 | geo | Geolocation information |
| 17 | source | Source |
| 18 | likes | Number of likes |
| 19 | quoted | Number of quotes |
| 20 | quote_url | Quote URL |
| 21 | refer_url | Refer URL |
| 22 | reply_url | Reply URL |
| 23 | photos | Photos |

The webpage corresponding to the Twitter tweet data field is as follows in Fig. 4,

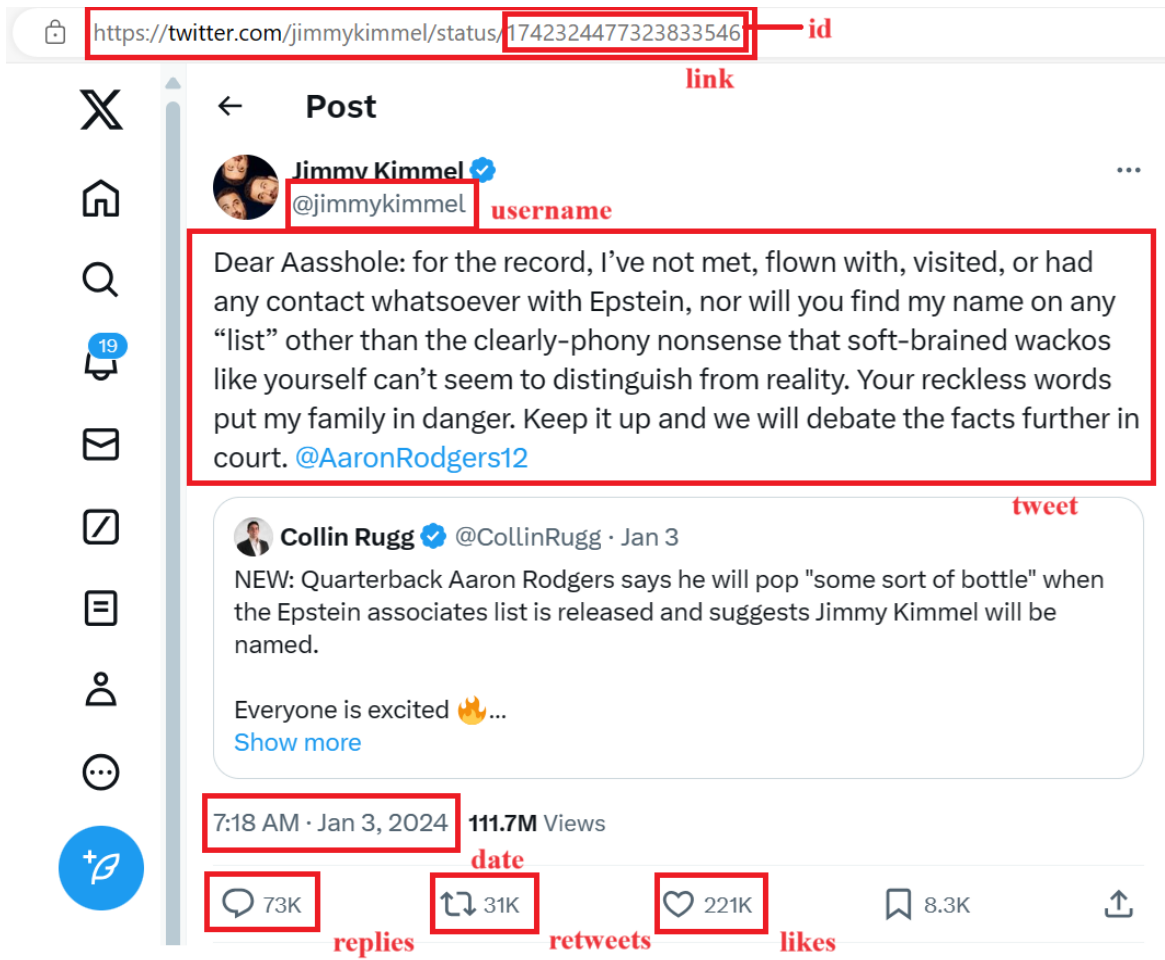

Fig. 4.

## 4.5.2.2. Comment data

The table and field explanations for Twitter comment data are as follows in Table 3,

Table 3

| NO. | Field | Field explanations |
| --- | --- | --- |
| 1 | post_id | Original tweet ID |
| 2 | comment_id | Comment ID |
| 3 | user_id | User ID |
| 4 | comment | Comment content |
| 5 | reply_to | Reply target (comment or user) |
| 6 | date | Comment date |
| 7 | source | Source |
| 8 | retweets | Number of retweets |
| 9 | likes | Number of likes |
| 10 | replies | Number of replies |
| 11 | mentions | Mentioned users |

| 12 | thread_id | Thread ID |
| 13 | reply_post_id | Replied tweet ID |

The webpage corresponding to the Twitter comment data field is as follows,

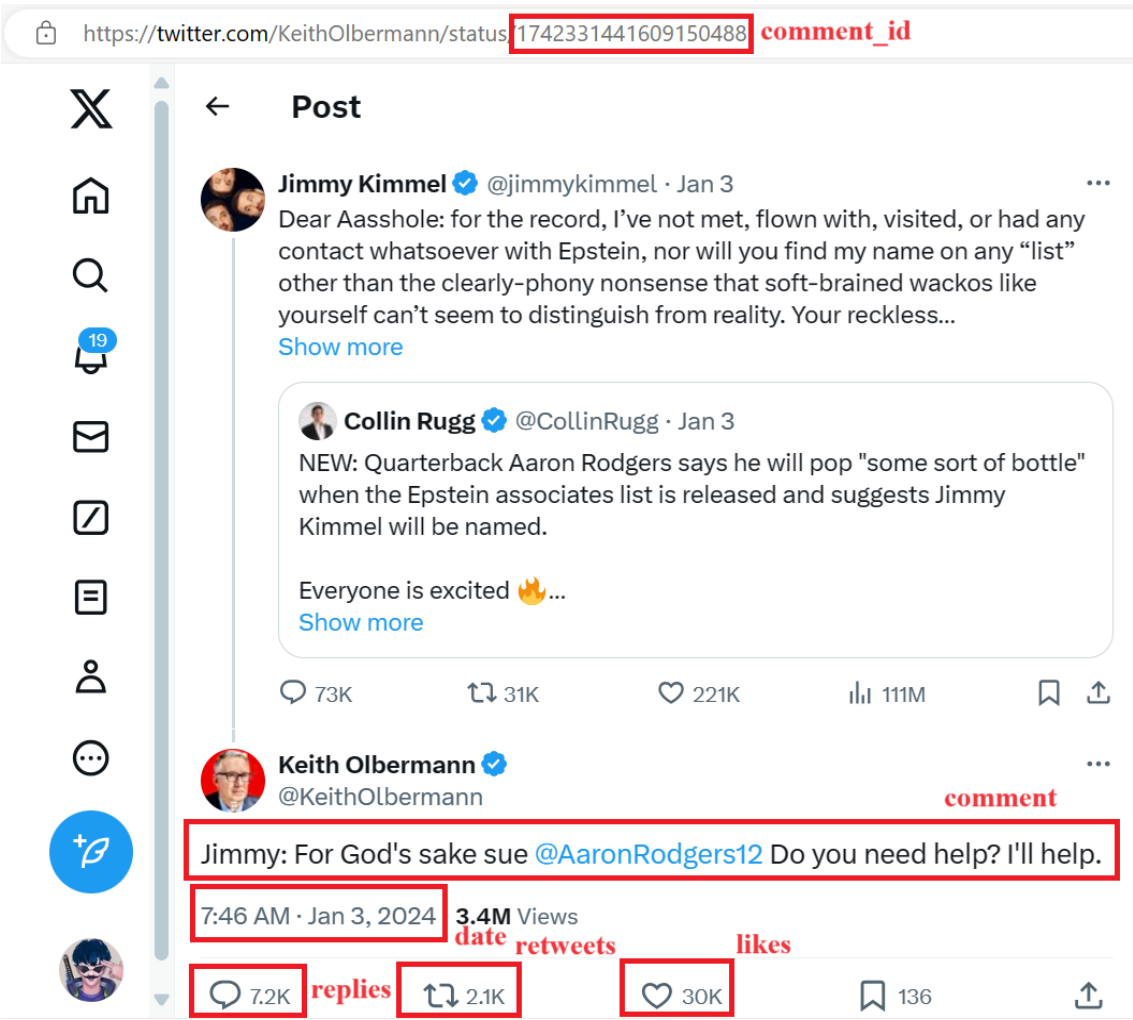

Fig. 5.

## 4.5.2.3. Repost data
The table and field explanations for Twitter repost data are as follows,

Table 4

| NO. | Field | Field explanations |
| --- | --- | --- |
| 1 | post_id | Original tweet ID |
| 2 | user_id | User ID |
| 3 | name | Full name |
| 4 | username | Username |

The webpage corresponding to the Twitter repost data field is as follows,

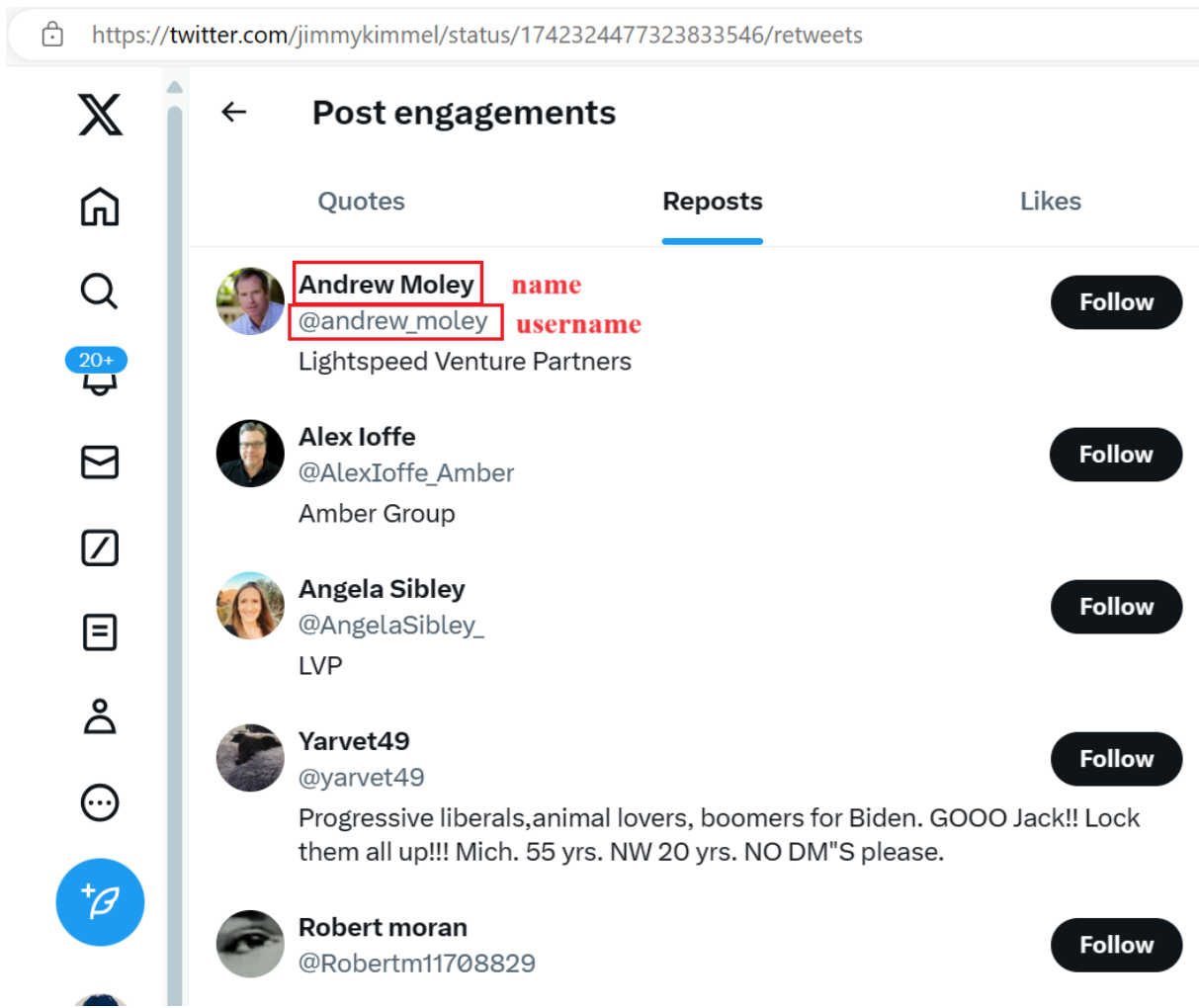

Fig. 6.

## 4.5.2.4. User data

The table and field explanations for Twitter user data are as follows,

Table 5

| NO. | Field | Field explanations |
|---|---|---|
| 1 | id | User ID |
| 2 | name | Full name |
| 3 | username | Username |
| 4 | bio | Biography |
| 5 | location | Location |
| 6 | url | Website URL |
| 7 | join_time | Join date |
| 8 | tweets | Number of tweets |
| 9 | following | Number of accounts being followed |
| 10 | followers | Number of followers |

| 11 | likes | Number of likes |
|---|---|---|
| 12 | media | Number of media items |
| 13 | private | Is private account |
| 14 | verified | Is verified account |
| 15 | profile_image_url | Profile image URL |
| 16 | background_image | Background image URL |

The web pages corresponding to Twitter user data fields are as follows,

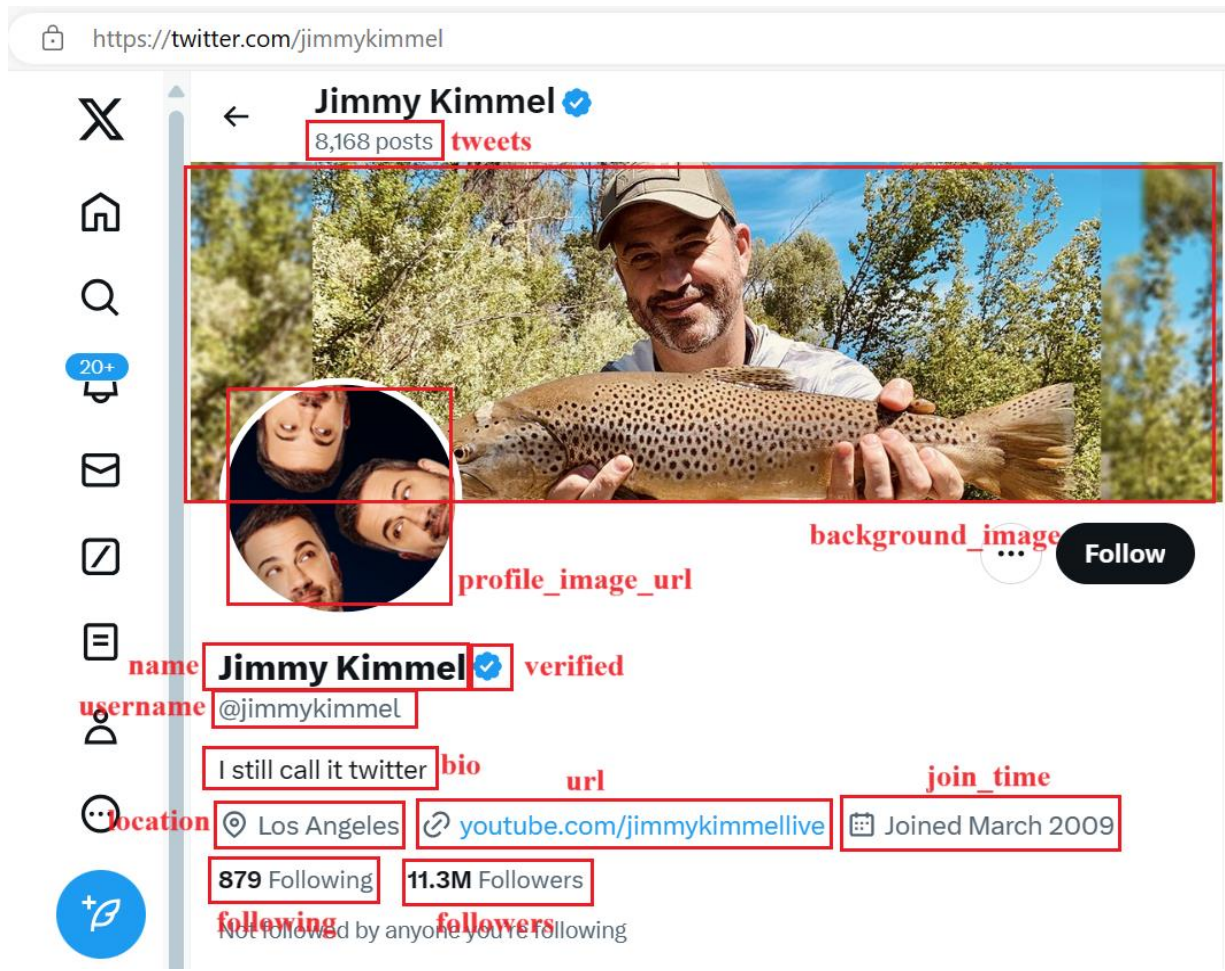

Fig. 7.

# 5. Proposed Method

As a researcher focusing on detecting rumors on social media, we use a two-pronged approach to refine the classification process. Firstly, we fine-tune the BERT model using a dataset of tweets and their subsequent comments from Twitter. This dataset will help the BERT model better understand the contextual nuances and linguistic subtleties in social media communication.

Secondly, we adopt the core concept of GraphSAGE and enhance it with an attention mechanism to meticulously aggregate edge attributes. This mechanism will be integrated into GraphSAGE trained on a dataset of interconnected tweets and comments, forming a comprehensive propagation graph. The aim is to improve the model's ability to classify tweets as either rumors or legitimate information.

To explain our methodology and the key components of the algorithm, we divide the exposition into two sections. The first section will outline the steps involved in fine-tuning the BERT model, focusing on the use of tweet-commentary sentences. The second section will provide a detailed explanation of the graph algorithm model, with particular emphasis on the attention-enriched aggregation of edge attributes and its role in rumor classification.

## 5.1. Enhancing BERT for Rumor Detection: Fine-Tuning with Twitter Data

In previous research on categorizing rumors on social networks, the detection of rumors is a challenging task that requires understanding the context and flow of information. Utilizing Twitter data for this purpose presents a unique opportunity due to its real-time nature and the abundance of user interactions. The Bidirectional Encoder Representations from Transformers (BERT) model, with its deep bidirectional understanding of language, is an ideal choice for such tasks. Fine-tuning BERT on Twitter data can enhance its ability to discern the nuances of social media communication and improve rumor detection capabilities.

### 5.1.1. Objectives and Benefits

The rationale for using Twitter data lies in its dynamic and interactive nature, which mirrors the propagation patterns of rumors. Tweets and their subsequent retweets or comments provide a sequential flow of information that can be effectively modeled by BERT's next sentence prediction (NSP) task. This task, as discussed in the seminal work by Devlin et al. (2018), allows BERT to learn the relationship between sentences, which is crucial for understanding the context in which rumors emerge and spread.

The fine-tuned BERT model, informed by NSP, is expected to exhibit an enhanced ability to identify rumors by recognizing patterns and contextual cues within the tweet discourse. This method is supported by the extant literature, which emphasizes the importance of contextual and sequential information in language understanding tasks, as demonstrated by the success of models such as ELMo and GPT.

Our proposed framework aims to contribute to the body of knowledge by providing a comprehensive, data-driven approach to rumor detection in social media. The advantages of employing this method for fine-tuning BERT are manifold. Firstly, it allows the model to learn from the sequential nature of social media conversations, which is crucial for understanding the context in which rumors may emerge. Secondly, the bidirectional context provided by BERT enables a more nuanced understanding of the semantic relationships within text, which is essential for distinguishing between genuine and fabricated information. Lastly, the fine-tuning process enhances the model's ability to adapt to the unique linguistic features and structures present in social media communications. In conclusion, the fine-tuned BERT model, through its improved understanding of social media conversations, offers a promising avenue for the development of more sophisticated algorithms capable of discerning truthful narratives from potential rumors with greater accuracy.

### 5.1.2. Sentence Pair Generation

Fine-tuning the BERT model using social media data involves creating sentence pairs that reflect the context and continuity found in platforms like Twitter. Inspired by the core idea of leveraging sentence relationships for model improvement from the research of Labruna et al.(2022) and Sun et al.(2019), we construct these pairs by concatenating tweets with their associated comments, or by treating a series of connected tweets and comments as a "conversation," mirrors the Next Sentence Prediction (NSP) task described in the research of Devlin et al.(2018). This task is designed to predict whether a given sentence is the actual follow-up to a preceding one, which is analogous to our method of using comments and forwarded tweets to establish sentence pairs. By focusing on the relationship between sentences, your method aligns with the NSP task's objective of capturing the contextual flow within a discourse, which is crucial for tasks that require understanding the broader context beyond individual sentences. Our fine-tuning approach leverages the inherent structure of Twitter conversations, where a tweet is followed by comments or retweets that can be considered the "next sentence".

In detail, we use two methods to combine tweets and their comments into the previous and next sentences required for fine-tuning the BERT model. The first method is to concatenate any commented tweet into its associated source tweet and use the comment of the commented tweet as the next sentence to form a sentence pair. The second method is to treat multiple tweets with continuous forwarding or comment relationships as "conversations", and then use any comment or forwarded tweet in the "conversations" as the next sentence, concatenating the consecutive tweets before the comment or forwarded tweet as the previous text. The pseudocode for these two methods is as follows.

**Input:**

$P$: *the set of post sets;*

$Model$: **Pre-trained BERT Model**

$Function\ pair\_generate1($ *source_tweet*, *input_comment* $)$

    *Initialize new list* $result\_list$

    *If* $input\_comment$ *has comments then*

        *Update* $previous\_sentence$ *by concatenating* $source\_tweet$ *and* $input\_comment$

        *For* $comment$ *in the comments of* $input\_comment$ *do*

            *Extend* $result\_list$ *with* $pair\_generate1($ *source_tweet*, *comment* $)$

        *End for*

    *Else*

        *Append the tuple of* $previous\_sentence$ *and* $input\_comment$ *into result_list*

    *End if*

    *Return* $result\_list$

$Function\ pair\_generate2($ *current_previous_sentence*, *input_comment* $)$

    *Initialize new list* $result\_list$

    *If* $input\_comment$ *has comments then*

        *Update* $current\_previous\_sentence$ *by concatenating* $current\_previous\_sentence$ *and* $input\_comment$

        *For* $comment$ *in the comments of* $input\_comment$ *do*

   *Extend* $result\_list$ *with* $pair\_generate2($ $current\_previous\_sentence$, $comment$ $)$

   *End for*

  *Else*

   *Append the tuple of* $current\_previous\_sentence$ *and* $input\_comment$ *into* result_list

  *End if*

  *Return* $result\_list$

*Alogorithm:*

*Begin*

  *Initialize new list* $sentence\_pair$

  *For* $\{p, p \in P\}$ *do*

   *For* $\{s, s \in p\}$ *do*

    *If* $s$ *is not a source tweet then*

     *Continue to the next tweet* $s$

    *End if*

    *Set* $previous\_sentence$ *as* $s$

    *Set* $candidate\_tweets$ *as the comments or other tweets that retweeted* $s$

    *For* $\{curr\_tweet, curr\_tweet \in candidate\_tweets\}$ *do*

     *Extend* $sentence\_pair$ *with the output of* $pair\_generate1$ *(* $previous\_sentence$, $input\_comment$ *)*

     *Extend* $sentence\_pair$ *with the output of* $pair\_generate2$ *(* $previous\_sentence$, $input\_comment$ *)*

    *End for*

   *End for*

  *End for*

  *Return* $sentence\_pair$

*Fine tune* $Model$ *with* $sentence\_pair$

*End*

**Output:**

$\mathcal{M}odel$: *BERT Model after fine-tuning*

*List:* $sentence\_pair$

### 5.1.3. Model Parameter Updates

The effectiveness of this fine-tuning strategy is underpinned by the layer-wise learning rate approach, which has been shown to mitigate catastrophic forgetting and enhance model performance on target tasks. This method, as detailed in Sun et al. (2019), allows for a more nuanced update of BERT's parameters, ensuring that the model does not overwrite its pretrained knowledge while still learning from the fine-tuning data. Specifically, we focus on fine-tuning the parameters of the 12th layer module of BERT. This decision is grounded in the understanding that higher layers of BERT are likely to capture more task-specific features, while lower layers retain general language understanding. By adjusting the 12th layer, we aim to retain the general pre-trained knowledge while adapting the model to the specifics of Twitter discourse.

In summary, the fine-tuning of BERT on Twitter data for rumor detection is a logical and data-driven approach. The use of NSP to mimic the flow of information on social media, coupled with a targeted fine-tuning strategy that focuses on the 12th layer module, is both academically sound and practically effective. This methodology not only enhances BERT's ability to understand the sequential nature of Twitter conversations but also provides a robust framework for the detection of rumors in social media platforms.

## 5.2. Graph Model Framework for Enhanced Rumor Detection

In this subsection, we present a concise model framework to facilitate readers' understanding of the proposed method. The primary objective of this study is to enhance rumor detection on social media by introducing an improved GraphSAGE algorithm that incorporates attention mechanisms for edge attributes. The enhanced Graph Neural Network, SAGEWithEdgeAttention, operates by first computing the feature differences between connected nodes to form edge attributes. These attributes, along with node features, are then processed through multiple graph convolutional layers with attention mechanisms. The attention mechanism allows the model to prioritize the most informative neighbors. By combining the node embeddings with the attention-weighted edge attributes, a more comprehensive representation of each node is generated. This representation is subsequently passed through fully connected layers to classify the central node, culminating in the final classification output. This method enables the model to better capture nuanced interactions within the propagation graph, thereby improving rumor detection performance.

To ensure a thorough understanding of our methodology, we first provide an in-depth explanation of the key components and underlying concepts of our algorithmic framework. Following this, we offer an integrated overview of the entire algorithm.

To illustrate our approach, we provide an example. Fig.8 (a) depicts the graph context of the target node (A) and its neighboring nodes (B, C, D, E, F). The model focuses on these connections to compute edge attributes and enhance node embeddings. Fig.8 (b) shows the structure of the SAGEWithEdgeAttention model, where node features and edge attributes are processed through the Edge Attention and Node Aggregation modules, resulting in a comprehensive node representation used for classification.

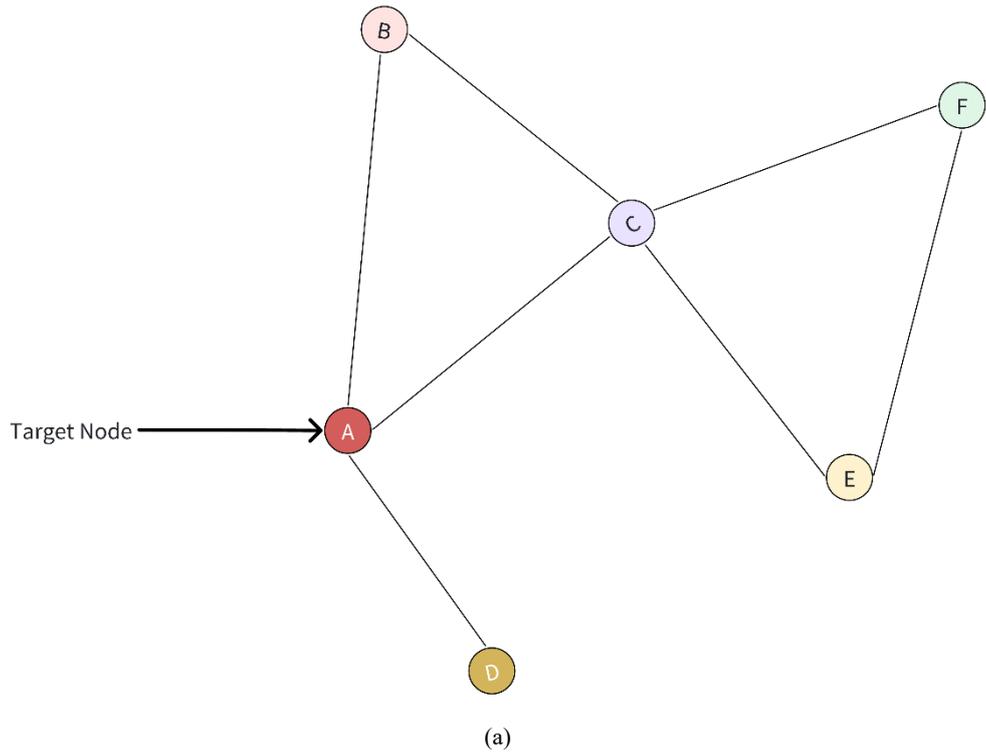

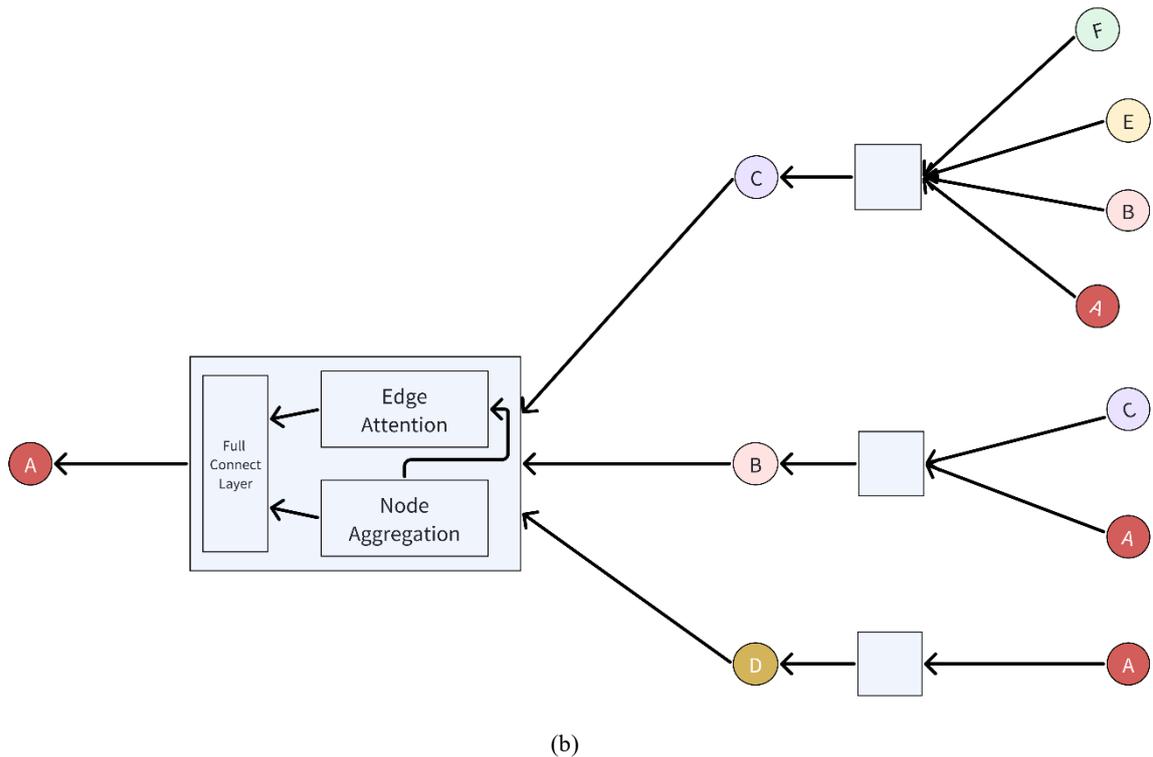

Fig.8: Overview of our model architecture with depth-2 convolutions (best viewed in color). (a) An example input graph with a small number of nodes. (b) The 2-layer neural network computes the embedding of node A by using the previous-layer representation of node A and its neighbors N(A) (nodes B, C, D). The Node Aggregation module calculates the properties of neighboring nodes for each node. Based on the aggregated node features, the Edge Attention module computes the weighted edge features. The embedding vector for node A is formed by concatenating the node aggregation result and the edge attention result.

## 5.2.1. Node Features Calculation

For effective rumor detection, the key content lies in the text of the posts. Traditional methods for text representation, such as word2vec (Mikolov et al., 2013) and GloVe (Pennington et al., 2014), provide static representations and struggle with polysemy. To address these limitations, we adopt BERT for extracting embeddings from post texts. BERT, a transformer-based model, has become a cornerstone in Natural Language Processing (NLP) due to its ability to capture contextual information. Unlike RNNs, BERT utilizes self-attention mechanisms to process sequences, allowing for more comprehensive text representations.

In our study, we utilize BERT's transformative architecture, which eschews the traditional RNN structure in favor of self-attention mechanisms. This enables BERT to capture a richer, more contextually informed representation of text. The fine-tuned BERT model, with its 12 layers and 768-dimensional hidden vectors, is further refined through our NSP-based fine-tuning process. We introduce a special token, [CLS], to the beginning of each sentence, and the final hidden state vector of this token is used to represent the entire sentence. This vector serves as the input feature for our interactive graph model, which is employed for the analysis of social media interactions. By integrating the NSP technique into our fine-tuning process, we augment BERT's ability to discern the intricate patterns of information flow on social media platforms. This method not only enhances the model's performance in rumor detection but also contributes to the broader field of NLP by demonstrating the effectiveness of bidirectional context modeling in understanding social media discourse. We user BERT model to produce the raw feature representation for a given tweet(i.e., node $v$), we use $r$ to donate the features of node $v$ generated by BERT, the calculation of process is donated as the formulation (1).

$$r_v = \{[r_v^1, r_v^2, \ldots, r_v^{768}] \leftarrow Bert(v), v \in V\} \tag{1}$$

Each tweet's text embedding vector is encapsulated as an attribute of its corresponding node. This attribute, along with the graph data, is subsequently integrated into the model.

### 5.2.2. Node Aggregation Module

Our proposed algorithm builds upon the GraphSAGE (SAmple and aggreGatE) framework introduced by Hamilton et al. (2017), which is an inductive framework developed to generate node embeddings for large graphs. The algorithm functions by learning a mechanism that produces embeddings by sampling and aggregating features from the local neighborhood of each node.

Building upon the foundational concept of GraphSAGE, our approach carefully aggregates the N-degree neighboring nodes associated with the source tweet within our framework. Each comment in our context serves as an indicator of the semantic content related to the source tweet or its antecedent tweet. Due to the importance of capturing the entire semantic landscape, we have chosen not to sample N-degree neighboring nodes, opting instead for a comprehensive aggregation to avoid omitting critical information.

In the original GraphSAGE, the aggregation of neighbor features is performed using a pooling operation, such as mean or max-pooling. However, in some literature, several works support the decision to forgo neighbor sampling in favor of a more exhaustive aggregation approach. Narayan et al. (2018) emphasizes the importance of preserving the structural integrity of the graph during the aggregation process. Additionally, Kipf and Welling (2017) provide a comprehensive analysis of the implications of neighbor sampling in GCN models, highlighting scenarios where complete aggregation may be preferable.

Building on the principles established in these seminal works, our methodology aims to provide a more holistic representation of the graph's semantic structure. This approach is particularly pertinent in the context of rumor detection, where capturing the nuances and subtleties within the propagation patterns of information is crucial.

In our study, we utilized the mean aggregator node attribute for aggregation, named as $SAGEConv$,

denoted symbolically as follows, donated as the following formulation (2).

$$\tag{2}$$

We define         as a collection of nodes draw from the set                                        .         denotes the current step in the outer loop and         denotes the node  's representation at this step: First, each node              aggregates the representations of the nodes in its first-degree neighborhood,                          , into a single vector         . This aggregation step depends on the representations generated at the previous iteration of the outer loop (i.e.,         ), and the              ("base case") representations are defined as the input node features. After aggregating the neighboring feature vectors, we sum the node's current representation         fed through a fully connected layer         and the elementwise mean of neighborhood vector         fed through a fully connected layer         , and this vector is activated by a nonlinear activation function σ, which transforms the representations to be used at the next step of the algorithm (i.e.,              ). The pseudocode for the Node Aggregation module is as follows.

*Function SAGEConv(*

    *X*<sub>(node features,</sub> $\{features\ of\ node\ v, v \in V\}$<sub>),</sub>

    *Relation*<sub>(Edge Relation,</sub> $\{(v,u), (v,u) \in E\}$<sub>)</sub>

*)*

    *Initialize weight matrices*

    *return* $agg\_res$

In conclusion, the SAGEConv function leverages node feature information to create embeddings that represent neighborhood nodes, which is particularly useful for applications involving evolving graphs or new subgraphs. Our proposed algorithm iteratively refines node representations. During each iteration, neighboring nodes initially share information through various types of relational paths. Subsequently, they update their representations by integrating the aggregated information from their neighbors along with their own. Consequently, the newly derived node representations encapsulate both textual and contextual information simultaneously.

### 5.2.3. Edge Features Calculation

In graph-structured data, edge attributes can capture the relationship between nodes. By using the difference between two text vectors as an edge attribute, one can model the relationship between texts more richly. This concept aligns with the practices in network theory and graph-based NLP models, where edge attributes are used to represent various types of relationships between nodes. For example, Jin et al. (2023) explores the integration of graph structures with neural networks to address complex tasks, particularly focusing on the nuanced representation of text differences through the contextual semantics of edge attributes in graph neural networks.

When considering the use of first-order differences between text vectors as edge attributes in a graph representation of text data, such as tweets and their replies, we draw inspiration from the field of sequential data analysis and natural language processing. Inouye et al. (2016) explores the importance of sequence order in set-based data, which can be analogous to capturing the sequential changes in textual data. The idea behind this paper inspires us to use sequence modeling in natural language processing, and utilizing first-order differences to represent the changes between texts can capture the dynamics of a conversation or the evolution of discourse. By treating the difference between text vectors as an edge attribute, we can model the progression and interaction between pieces of text more effectively. This process can be expressed by the following formulation (3).

$$r_{vu} = \{[r_{vu}^1, r_{vu}^2, \ldots, r_{vu}^{768}] \leftarrow r_v - r_u, (v, u) \in E\} \quad (3)$$

### 5.2.4. Edge Attention Module

In the realm of deep learning and graph neural networks (GNNs), the attention mechanism has emerged as a crucial technique for handling data with variable lengths, such as textual inputs in machine translation tasks. This mechanism enables models to dynamically focus on relevant parts of the input, thereby enhancing predictive accuracy (Vaswani et al., 2017; Bahdanau et al., 2014). By directing computational resources towards pertinent input components, the attention mechanism improves performance across tasks requiring nuanced data understanding.

Integrating edge attributes into the attention mechanism represents a significant advancement within GNNs. This integration, as explored by Wu et al., enables a more nuanced understanding of node relationships in graphs. Lin et al. demonstrated that this approach enhances representation learning by incorporating the broader social context, focusing on posts that semantically infer the target claim and outperforming existing methods.

The incorporation of attention mechanisms with edge attributes allows for a detailed analysis of node relationships, crucial for applications like rumor detection on social networks. Edge attributes capture interaction nuances such as temporal sequence, intensity, and sentiment, enriching the understanding of content propagation patterns and veracity. By integrating these attributes into the attention mechanism, GNNs can differentiate and prioritize interactions based on their significance, recognizing that not all interactions equally influence classification outcomes. For example, rapidly disputed posts are less likely to propagate rumors compared to uniformly endorsed ones. This capability underscores the importance of leveraging edge attributes within the attention mechanism to discern complex interactions inherent in online social media discourse.

To effectively capture edge attribute representations, we employ an attention mechanism to compute $att$. The pseudocode for the attention module is outlined as follows,

*Function att(*

    *$X$(input features of nodes),*

    *$J$ (number of attention heads),*

    *$d_X$(feature dimension of $X$)*

*)*

    *Initialize weight matrices $W_Q$(Query Matrix)*

    *Initialize weight matrices $W_K$（Key Matrix）*

    *Initialize weight matrices $W_V$（Value Matrix）*

*Initialize weight matrices* $W_O$ *(Matrix for linear transformations),*

*#Linear transformations*

$Q = X \cdot W_Q$

$K = X \cdot W_K$

$V = X \cdot W_V$

*#Split into multiple heads*

$Q = split(Q, H)$

$K = split(K, H)$

$V = split(V, H)$

*#Calculate attention scores*

$Attention\_scores = \dfrac{K \cdot Q^T}{\sqrt{d_X/J}}$

$Attention\_probabilities = Softmax(Attention\_scores)$

*#Apply attention weights to values*

$Attention\_out = Attention\_probabilities \cdot V$

*#Output*

$Concatenated\_output = Concat(Attention\_output_1, \ldots, Attention\_output_h)$

$Final\_output = Concatenated\_output \cdot W_O$

*return* $Final\_output$

This main strategy is described as follows,

1. Edge attention score. For each node $v \in V$, and each user neighbor $u \in N_v$, we calculates an attention weight $e_{vu}$ by using an attention function $att^k$ on node aggregation matrix $h_v^k$ and $h_u^k$ of the current layer $k$ for both nodes $\{(v, u), (v \in V, u \in N_{(v)})\}$, donated as the following formulation (4).

$$e_{vu}^k = \{att^k(concat(h_v^k, h_u^k)), (v \in V, u \in N_{(v)})\} \tag{4}$$

2. Attention scores normalization. We then normalize scores using a SoftMax function to get a probability of each scores, donated as the following formulation (5).

$$a_{vu}^k = Softmax(e_{vu}^k) = \{\dfrac{exp(e_{vu_i}^k)}{\sum exp(e_{vu_i}^k)}, u_i \in N(v)\} \tag{5}$$

3. Edge features calculation. The normalized attention scores are then used to compute the edge features for each node, donated as formulation (6).

$$\{MEAN_v(Softmax(MEAN(a_{vu}^k)) \cdot r_{vu}), v \in V\} \tag{6}$$

First, we compute the average normalized attention score for each node $v$, donated as the following formulation (7).

$$attention\_scores = Softmax(MEAN(a_{vu}^k)) \tag{7}$$

Then we multiply edge features $r_{vu}$ and average normalized attention score at the element level and compute the mean of the weighted edge attributes for each node in the graph, donated as the following formulation (8).

$$\{MEAN_v(r_{vu} \cdot attention\_scores), (v \in V\ u \in N_{(v)})\} \tag{8}$$

The pseudocode for the above steps is as follows,

*Function propagate_att(*

    *Relation*(Edge Relation, $\{(v,u), (v,u) \in E\}$),

    *X*(node features, $\{features\ of\ node\ v, v \in V\}$),

    *edge_attr*(node features, $\{features\ of\ edge\ (v,u), (v,u) \in E\}$),

    *attention$^k$*(attention function, $\{k, k \in (1\ to\ K)\}$)

*)*

    $edge\_attention = \{attention^k(Concat(r_v, r_u)), (v \in V, u \in N_{(v)})\}$

    $weighted\_edge\_attr = Softmax(MEAN(edge\_attr)) \cdot edge\_attr$

    *return* $\{MEAN_v(weighted\_edge\_attr), v \in V\}$

## 5.2.5. SAGEWithEdgeAttention

The SAGEWithEdgeAttention algorithm, an extension of the GraphSAGE framework, integrates an attention mechanism to more effectively aggregate edge attributes. These attributes are derived from the first-order differences between the features of source and destination nodes, thereby enhancing the model's capacity to encapsulate the relational dynamics within the network. The attention mechanism is seamlessly integrated within the message-passing paradigm, allowing nodes to prioritize information from neighboring nodes based on the computed attention scores.

The algorithm iteratively conducts a feature aggregation process over K iterations. This process includes utilizing the SAGEConv layer followed by a ReLU activation function and a Dropout layer for regularization. Within the *propagate_att* function, the attention mechanism is applied to the aggregated features and edge attributes. The result of this attention-weighted aggregation is a concatenation of enhanced node features and edge attributes, which undergo further processing through a linear layer.

The accompanying pseudocode delineates the operational framework of the SAGEWithEdgeAttention algorithm. This representation incorporates several functions previously defined within the exposition, node features calculation, node aggregation module, edge features calculation, edge attention module.

**Algorithm: SAGEWithEdgeAttention**

*Input:*

*Graph:* $G(V, E)$;

*node features:* $\{r_v, v \in V\}$;

*depth:* $K$;

**Output:**

*Node category distribution vector:* $\{z_v, v \in V\}$

**Algorithm:**

Begin

    *Initialize weight matrices:* $W_1, W_2, W_3, W_4$

    *Initialize SAGE aggregation module:* $\{SAGEConv^k, k \in (1\ to\ K)\}$

    *Initialize Edge Attention module:* $\{att^k, k \in (1\ to\ K)\}$

    *#Compute edge attributes*

    $r_{vu} = \{r_v - r_u, (v \in V, u \in N_{(v)})\}$

    *#Apply linear transformation to edge attributes*

    $l_1 = ReLU(r_{vu} \cdot W_1 + b_1)$

    $edge\_attr = l_1 \cdot W_2 + b_2$

    *for* $k \in (1\ to\ K)$ *do*

        *if k=1 then*

            $h_v^{k-1} = r_v$

        *else*

            $h_v^{k-1} = h_v^k$

        *End if*

        *#Apply node features aggregation module*

        $h_v^k = \{DropOut(RuLE(SAGEConv^k(h_v^{k-1}))), v \in V\}$

        *#Apply edge attention mechanism*

        $edge\_att = \{propagate\_att(\{(v,u), u \in N_{(v)}\}, h_v^k, edge\_attr, att^k), v \in V\}$

        $h_v^k = \{W_3 \cdot Concat(h_v^k, edge\_att), v \in V\}$

    *End for*

    *return* $Output output = W_4 \cdot RuLE(h_v^k)$

*End*

In conclusion, by incorporating an attention mechanism, SAGEWithEdgeAttention augments GraphSAGE's capability to process node features and edge attributes dynamically. This innovation enables a nuanced weighing of neighboring nodes' contributions during the feature aggregation phase, which may significantly bolster the model's performance in tasks such as rumor detection on social networks, where the accurate representation of relational dynamics is paramount for discerning propagation patterns.

# 6. Experiments

The primary objective of our experimental process is to evaluate the effectiveness of the proposed methodology in enhancing rumor detection on social media. By incorporating advanced techniques such as BERT for text representation and the SAGEWithEdgeAttention algorithm for graph-based analysis, we aim to improve the model's ability to accurately classify social media content. Our experiments are designed to assess the performance of these techniques under various conditions, ensuring a comprehensive understanding of their impact. The subsequent sections will detail the preprocessing steps, dataset construction, evaluation metrics, parameter settings, and model training procedures, providing a thorough overview of the experimental framework and its implementation.

## 6.1. Preprocessing

### 6.1.1. Dataset for fine-tuning

Within the purview of BERT model fine-tuning, our approach meticulously employs the previously delineated sentence pair generation technique to curate the textual data. To this end, the sentence pairs, which exhibit a relational connectivity between the preceding and subsequent segments, are designated as affirmative instances. Subsequently, an adversarial instance is constructed by amalgamating the preceding sentence from the affirmative pair with the subsequent sentence derived from an alternate set of sentence pairs. These latter sentence pairs are procured via a stochastic sampling mechanism from a corpus of source tweets. Furthermore, to enhance the robustness of the training regimen, the ratio of affirmative to adversarial instances is meticulously balanced at a calibrated ratio of 1:5. This nuanced methodology is engineered to simulate the complex interplay of linguistic sequences, thereby fortifying the BERT model's discernment capabilities between coherent and non-coherent textual segues.

### 6.1.2. Dataset for SAGEWithEdgeAttention

For experimental evaluation, we enhance the static graph representation of each source tweet by adding new nodes and edges. These are derived from responsive tweets generated by the source tweet at six-hour intervals, thereby introducing temporal dynamics to the graph. Fig.9 illustrates the difference between dynamic and static news propagation networks. As shown in Fig.9 (left), the news dissemination graph evolves over time in the dynamic graph, with users' spreading behaviors occurring at timepoints t1, t2, t3, and t4. In contrast, Fig.9 (right) depicts a static graph that captures the graph structure without continuous temporal dynamics.

Consistent with the definition of dynamic graphs established in the preceding text, we extend this approach by chronologically ordering the sequence of static graphs associated with each source tweet. This methodological enhancement results in the construction of dynamic graphs that encapsulate the temporal evolution of the network in response to each source tweet.

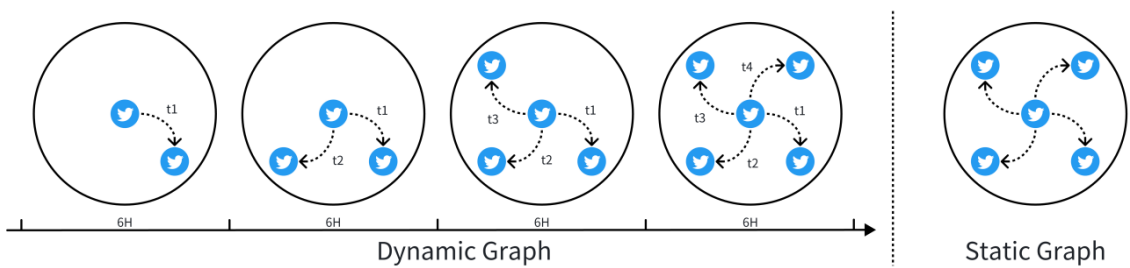

Fig. 9. Left: Dynamic Graph representing the propagation of a piece of social media news. Each Twitter bird icon denotes a tweet, and each arrow represents a share or retweet, accompanied by a timestamp. Right: Static Graph representing the propagation of the same piece of social media news. Here, each arrow indicates a share or retweet, but without a timestamp.

## 6.2. Experimental Setup

In this section, we meticulously outline the evaluation metrics, including accuracy, which is our primary metric of interest, despite the challenges it may present due to class imbalance.

### 6.2.1. Evaluation metrics

The algorithms selected for comparative analysis are chosen based on their relevance and performance in similar tasks within the domain of social network analysis. We then proceed to elaborate on the process of parameter selection, which is crucial for the fine-tuning of our graph neural network model.

Accuracy serves as a foundational metric for assessing the classification model's performance by measuring the ratio of correct predictions (true positives and true negatives) to the total number of evaluated cases. However, the presence of imbalanced class distributions can distort the interpretation of accuracy. Models that predominantly classify examples into the majority class may falsely appear to perform well. To address this issue, our study utilizes the F1 score to comprehensively evaluate the model's performance across all classification tasks. Additionally, we employ accuracy, Micro F1, and Macro F1 as supplementary metrics to provide a holistic assessment of model performance. The formulas for these metrics are defined as follows,

$$Precision = \frac{TP}{TP + FP} \tag{9}$$

$$Recall = \frac{TP}{TP + FN} \tag{10}$$

$$F1_c = 2 * \frac{Precision_c * Recall_c}{Precision_c + Recall_c}, c \in C \tag{11}$$

$$Accuracy = \frac{1}{N} \sum_{i=1}^{N} \delta \left( \hat{y}_i == y_i \right) \tag{12}$$

$$Micro\_F1 = \frac{2 * TP}{2 * TP + FP + FN} \tag{13}$$

$$Macro\_F1 = \frac{1}{Number\ of\ C} \sum_{c \in C} F1_c \tag{14}$$

Where, $TP$ is the total number of true positive instances across all classes. $FP$ is the total number of false positive instances across all classes. $FN$ is the total number of false negative instances across all classes. $TN$ is the total number of true negative instances across all classes. $C$ is the set of all distinct labels or categories in the classification task. $c$ stands for an individual label or category to which a sample belongs. $N$ is the total number of instances or samples. $y_i$ is the true label of the i-th instance. $\hat{y}_i$ is the predicted label for the i-th instance by the classifier. $\delta$ is the Kronecker delta function, which returns 1 if the argument is true (i.e., the predicted label equals the true label) and 0 otherwise.

### 6.2.2. Parameters settings

When implementing SAGEWithEdgeAttention to learn node representations in the propagation graph, critical parameters are carefully selected to enhance model performance and generalization. The *in_channels* parameter is set to 756, matching the dimensionality of the input feature space, which encapsulates the inherent characteristics of the nodes within the graph. To enable the model to capture complex patterns, a number of *hidden_channels* are defined as 64, representing the dimensionality of the hidden layers where intermediate feature transformations occur. The output layer, represented by *out_channels*, is set to 5, aligning with the number of distinct classification labels present in the dataset.

Introducing *num_heads* as 2 in the attention mechanism enables the model to concurrently process diverse feature representations, potentially enriching node representations with greater depth and detail. Additionally, to mitigate risks of overfitting, a dropout ratio of 0.5 is applied. This technique randomly deactivates a fraction of input units during each training update, encouraging the network to learn more resilient features.

For the optimization of the model's parameters, the AdamW optimizer is utilized, with a *learning_rate*, set to 0.01 that dictates the size of the steps taken towards minimizing the loss function. This adaptive learning rate method, combined with weight decay encapsulated in AdamW, ensures that the model converges efficiently while maintaining a focus on generalization.

### 6.2.3. Model training

The loss function utilized in our proposed algorithm comprises the cross-entropy loss, which measures the disparity between the predicted probability distributions and the ground truth.

$$Loss = -\sum_{n=1}^{N} \sum_{c \in C} y_c(x_n) * log(\widehat{y_c}(x_n))$$

(15)

$C$ is the set of all distinct labels or categories in the classification task. $c$ stands for an individual label or category to which a sample belongs. $N$ is the total number of instances or samples. To enhance the performance of our model, we employ an iterative process across all training examples in each epoch, persisting until the loss value, as determined by the above Equation, either stabilizes or the pre-set maximum number of epochs is exhausted. Throughout this training regimen, we leverage the back-propagation algorithm to refine all trainable parameters (Goller & Kuchler, 1996).

Various optimizers are available for training neural network-based models, including SGD, AdaGrad, RMSProp, and Adam (Yu, 2020; Kingma & Ba, 2014; Shao, Pi, & Shao, 2019). While SGD and its variants like batch gradient descent and mini-batch gradient descent are widely employed, they come with inherent limitations: 1) selecting an appropriate initial learning rate can be challenging, 2) parameters are updated uniformly with a single learning rate, which may not be optimal given the varying importance and frequency of parameters, particularly in sparse data scenarios, and 3) these algorithms can struggle to escape numerous suboptimal local minima and saddle points.

To address these challenges, recent advances have introduced optimization algorithms with adaptive learning rate adjustments. For instance, AdaGrad customizes the learning rate for each parameter, applying smaller updates to frequently occurring features and larger updates to less common ones. Additionally, momentum-based optimization algorithms like RMSProp have been developed to mitigate issues of oscillations and slow convergence. By utilizing the exponential moving average of squared gradients, RMSProp adjusts the learning rate dynamically, thereby enhancing convergence speed and alleviating problems associated with saddle points and local minima.

AdamW, an extension of the Adam optimizer, combines the benefits of AdaGrad and RMSProp while addressing the problem of weight decay. By directly incorporating a weight decay term into the optimization process, AdamW enhances generalization and performance in practical applications (Loshchilov & Hutter, 2019). By leveraging both the first and second moment estimates of gradients, AdamW dynamically adjusts the learning rate for each parameter, effectively navigating potential suboptimal local minima. This makes AdamW particularly well-suited for our model, which frequently handles high-dimensional and sparse data.

In our research, we utilize the method of AdamW optimizer to streamline the training process, benefiting from its adaptive learning rate capabilities and resilience against common optimization challenges.

## 6.3. Results and Analysis

To fine-tune the BERT model and train SAGEWithEdgeAttention, we partitioned the dataset into 70% for training, 20% for validation, and 10% for testing. The optimal model selection process is as follows: at the beginning of each epoch, the training set is randomly divided into batches, which are sequentially fed into the model for parameter updates via backpropagation to minimize the loss. The model's performance is evaluated using the validation set at the end of each epoch. Training continues until no improvement in validation loss is observed for 50 consecutive epochs. Following training completion, we evaluate the optimal model's performance on the testing set. Figures 9 (a), (b), (c), and (d) depict the evolution of accuracy and loss for our proposed method, SAGEWithEdgeAttention, across training and validation sets as epochs progress. The x-axis represents the number of epochs, the y-axis on the left indicates loss, and the y-axis on the right indicates accuracy. The dotted line on the x-axis denotes the epoch count where the model achieves optimal performance.

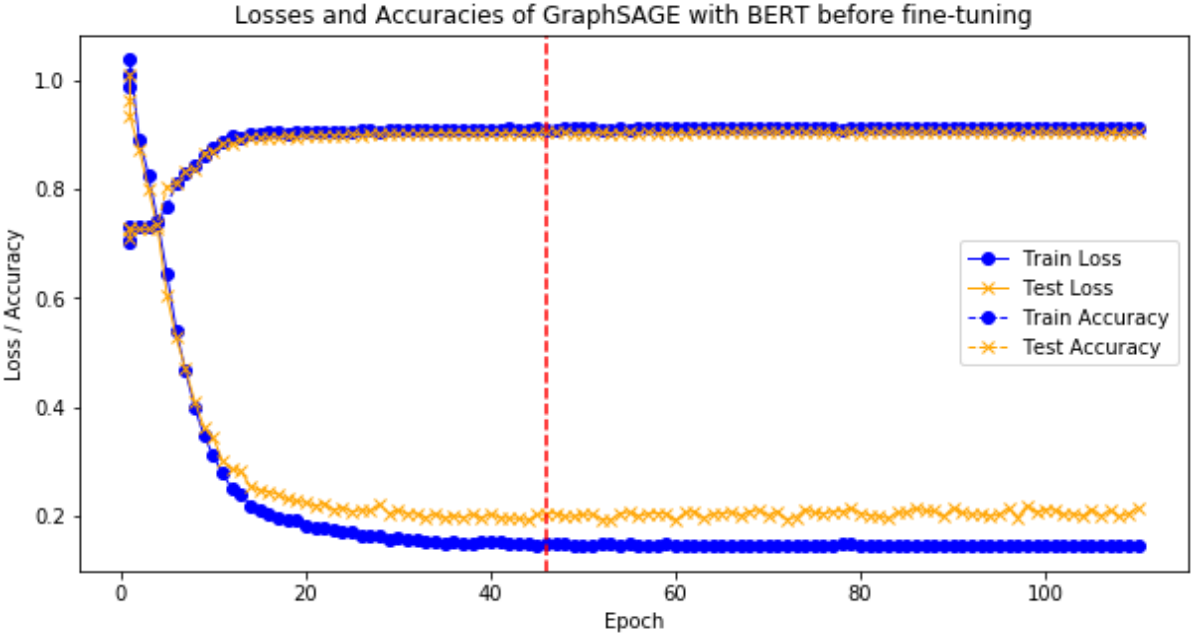

(a)

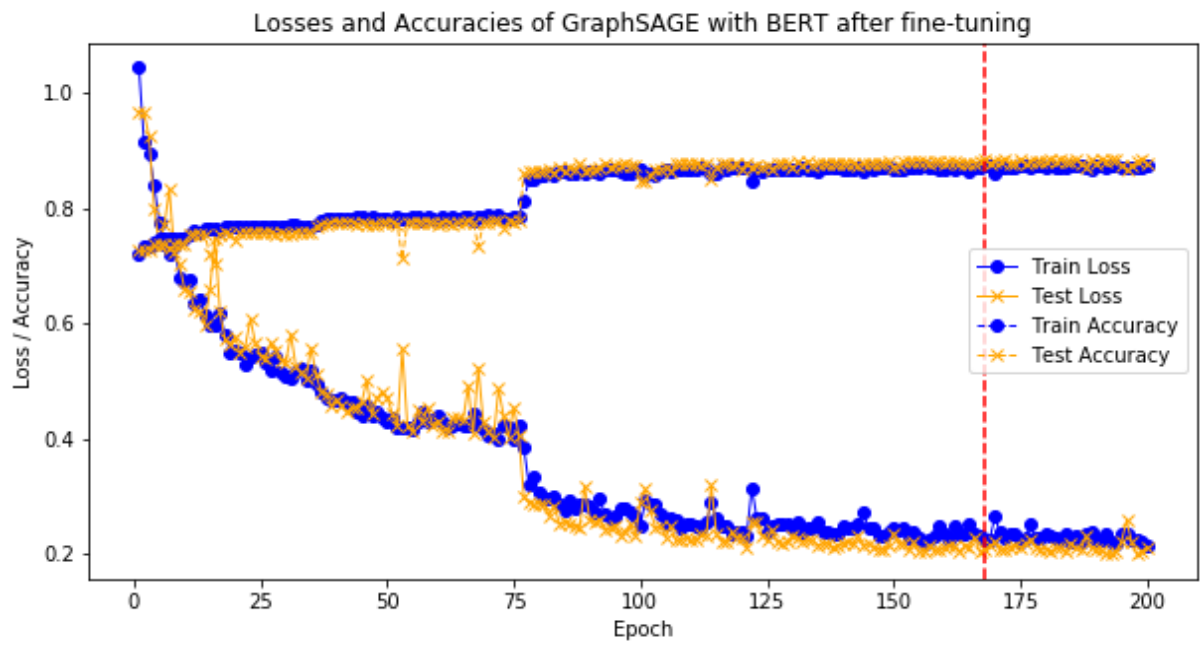

(b)

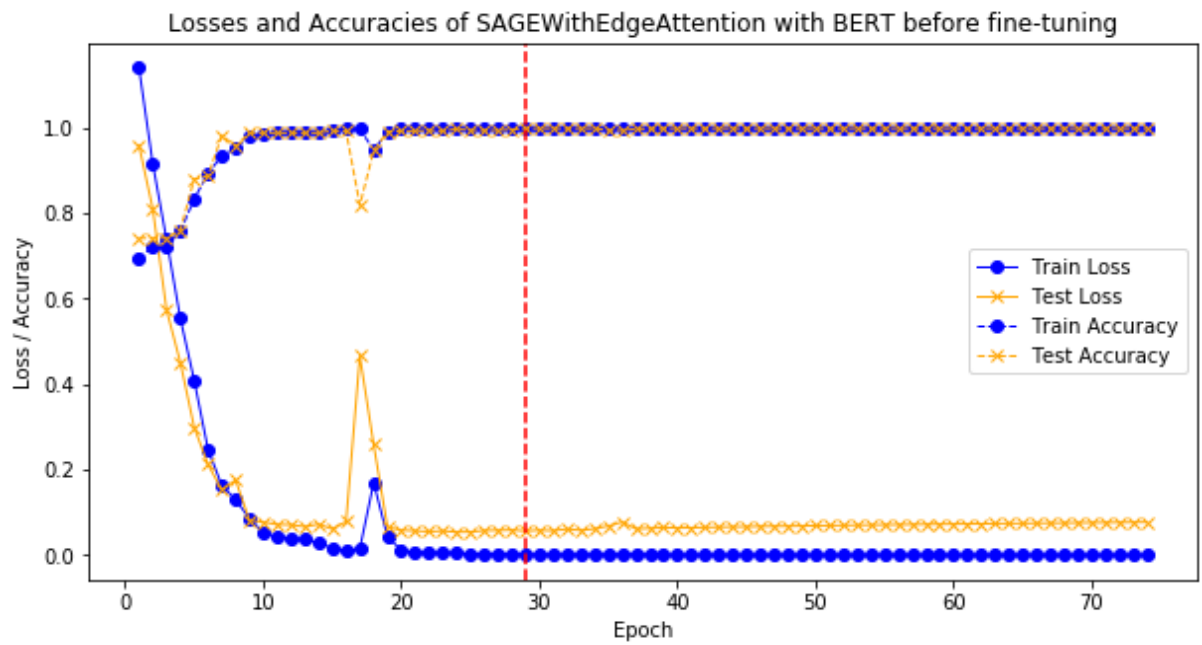

(c)

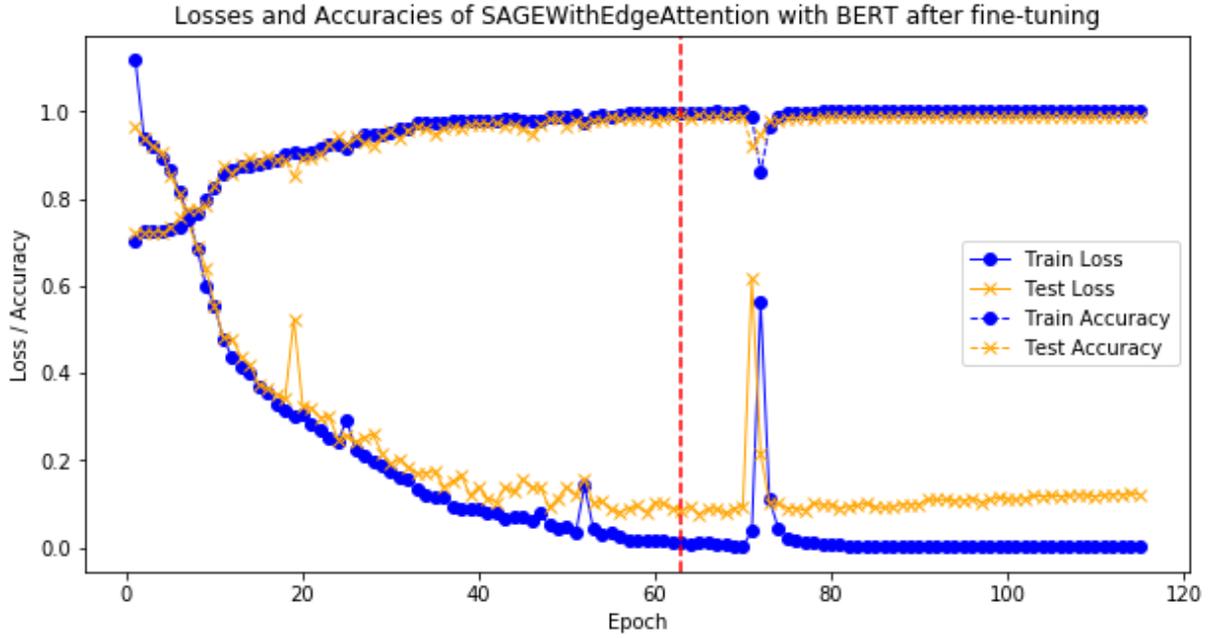

(d)

Fig. 10. The selection of the optimal model

The following table provided delineates the comparative performance of graphSAGE and SAGEWithEdgeAttention, in the context of leveraging BERT-generated word vectors, both before and after fine-tuning.

Table 6

Performance comparison for rumor classification

| Algorithm | Accuracy | Micro F1 | Macro F1 | F1 False | F1 Pants on Fire | F1 Half True | F1 Mostly False | F1 Mostly True |
|---|---|---|---|---|---|---|---|---|
| graphSAGE with BERT berfore fine-tuning | 0.7139 | 0.7227 | 0.2210 | 0.8318 | 0.0343 | 0.0 | 0.2389 | 0.0 |
| SAGEWithEdgeAttention with BERT berfore fine-tuning | 0.7227 | 0.7227 | 0.2637 | 0.8394 | 0.2705 | 0.0 | 0.2087 | 0.0 |
| graphSAGE with BERT after fine-tuning | 0.8813 | 0.8813 | 0.5670 | 0.9311 | 0.9323 | 0.9715 | 0..0 | 0.0 |
| SAGEWithEdgeAttention with BERT after fine-tuning | **0.9974** | **0.9974** | **0.9960** | **0.9988** | **0.9894** | **1.0** | **0.9975** | **0.9944** |

The results are as follows, before fine-tuning the BERT model, graphSAGE achieved a accuracy score of 0.7139. Post fine-tuning, this score improved to 0.8813. SAGEWithEdgeAttention demonstrated a superior performance before fine-tuning, with a score of 0.7227. After fine-tuning the BERT model, the score marginally increased to 0.9974. From the table, it is evident that the SAGEWithEdgeAttention model with BERT after fine-tuning, achieved the highest F1 scores across all five label categories. The empirical analysis reveals that both graph-based algorithms exhibit an enhancement in performance metrics when utilizing word vectors derived from a fine-tuned BERT model as opposed to a pre-fined-tuned one. Notably, the SAGEWithEdgeAttention algorithm, which integrates attention mechanisms to weigh the importance of different edges in the graph, has demonstrated a more pronounced improvement, suggesting a heightened sensitivity to the nuanced features captured through fine-tuning. The graphSAGE algorithm, while also benefiting from fine-tuning, exhibited a more modest increase, indicating a potentially less pronounced reliance on the fine-grained adjustments afforded by BERT's fine-tuning process.

# 7. Conclusion and future work

In the context of global elections with near half of the world's population participating, rumor detection has become a crucial task for maintaining information transparency and public trust. This study constructs a real-world dataset of politically related rumors by collecting data from the PolitiFact website and Twitter. It proposes an effective rumor identification method and builds an intelligent classifier based on this method to distinguish source tweets into five categories, "False", "Pants on Fire", "Half True", "Mostly False" or "Mostly True", achieving effective discrimination of rumors. The two-stage technical route to achieve this goal includes,

We fine-tune the BERT model and use a dataset containing tweets and their corresponding comments as training material to enhance the model's ability to understand language features and subtle contextual differences specific to the social media domain. Through this process, an initial feature vector rich in contextual information is generated for each tweet node.

We introduce a novel graph neural network, SAGEWithEdgeAttention, as an extension of the GraphSAGE model. This algorithm integrates an attention mechanism to optimize the aggregation process of edge attributes. The specific operation involves first constructing a graph with tweets and their corresponding comments as edges, and then inputting it into the algorithm as input, and outputting the category of the source tweet corresponding to the graph data. By using the first-order difference between the source node and terminal node features as edge attributes, the addition of the attention mechanism allows the model to perform weighted fusion of neighbor node information based on importance, improving the targeting and efficiency of feature aggregation.

Through experiments conducted on our constructed real-world Twitter dataset, our proposed SAGEWithEdgeAttention algorithm demonstrated significant performance improvements over the graphSAGE baseline in the task of rumor detection. In our comprehensive comparative analysis, we found that when utilizing word embeddings derived from the fine-tuned BERT model, both graph-based algorithms achieved an enhancement in performance metrics. Notably, the SAGEWithEdgeAttention algorithm, by integrating an attention mechanism to assess the importance of different edges in the graph, exhibited even more remarkable improvements, indicating its higher sensitivity to subtle features captured during fine-tuning.

In summary, we combine the deep contextual understanding ability of the pre-trained model BERT with the latest advances in graph neural network technology, especially the SAGE variant with integrated attention mechanism, to provide a theoretically rigorous and practically efficient solution for rumor detection tasks in social networks. Compared with traditional methods that rely solely on content analysis or models that only consider the time dimension, our method shows significant performance improvement in practical applications, highlighting the advancement and effectiveness of capturing and parsing propagation patterns in complex information environments.

In future research, we plan to focus on integrating more sophistic social network data of users, including follower-followee relationships within the community and records of common participation between users. These social relationships essentially constitute a rich graph structure that will provide us with valuable resources for deepening propagation graph analysis. It is worth noting that although most current rumor detection algorithms mainly rely on supervised learning frameworks, data in real-world scenarios are often unlabeled, which is a major challenge we face. In view of this, we plan to introduce transfer learning methods to address this challenge and perform rumor detection tasks more effectively.